\documentclass[twocolumn]{aastex631}

\newcommand{\swift}{\emph{Swift}}
\newcommand{\wise}{\emph{WISE}}

\shorttitle{The Type Icn SN\,2021csp}
\shortauthors{Perley et al.}

\begin{document}

\title{The Type Icn SN\,2021csp:  \\
Implications for the Origins of the Fastest Supernovae and the Fates of Wolf-Rayet Stars}

\newcommand{\okc}{The Oskar Klein Centre, Department of Astronomy, Stockholm University,  AlbaNova, SE-106 91 Stockholm, Sweden}
\newcommand{\okcp}{The Oskar Klein Centre, Department of Physics, Stockholm University,  AlbaNova, SE-106 91 Stockholm, Sweden}
\newcommand{\caltech}{Division of Physics, Mathematics, and Astronomy, California Institute of Technology, Pasadena, CA 91125, USA}
\newcommand{\weizmann}{Department of Particle Physics and Astrophysics, Weizmann Institute of Science, 76100 Rehovot, Israel}
\newcommand{\berkeley}{Department of Astronomy, University of California, Berkeley, CA 94720-3411, USA}
\newcommand{\coo}{Caltech Optical Observatories, California Institute of Technology, Pasadena, CA 91125, USA}
\newcommand{\millerfellow}{Miller Institute for Basic Research in Science, University of California, Berkeley, CA 94720, USA}
\newcommand{\eso}{European Organisation for Astronomical Research in the Southern Hemisphere (ESO), Karl-Schwarzschild-Str. 2, 85748 Garching b.\ M{\"u}nchen, Germany}
\newcommand{\esochile}{European Organisation for Astronomical Research in the Southern Hemisphere (ESO), Vitacura Alonso de C\'ordova 3107 Vitacura, Casilla 19001 Santiago de Chile, Chile}
\newcommand{\fsu}{Department of Physics, Florida State University, Tallahassee, FL 32306-4350, USA}
\newcommand{\sheffield}{Department of Physics and Astronomy, University of Sheffield, Hicks Building, Hounsfield Road, Sheffield S3 7RH, UK}
\newcommand{\lbnl}{Physics Division, Lawrence Berkeley National Laboratory, 1 Cyclotron Road, Berkeley, CA 94720, USA}
\newcommand{\uca}{Universit\'e Clermont Auvergne, CNRS/IN2P3, LPC, F-63000 Clermont-Ferrand, France}
\newcommand{\ipac}{IPAC, California Institute of Technology, 1200 E. California Blvd, Pasadena, CA 91125, USA}
\newcommand{\dirac}{DIRAC Institute, Department of Astronomy, University of Washington, 3910 15th Avenue NE, Seattle, WA 98195, USA}
\newcommand{\minnesota}{School of Physics and Astronomy, University of Minnesota, Minneapolis, MN 55455, USA}

\author[0000-0001-8472-1996]{Daniel A. Perley}
\email{d.a.perley@ljmu.ac.uk}
\affiliation{Astrophysics Research Institute, Liverpool John Moores University, \\ IC2, Liverpool Science Park, 146 Brownlow Hill, Liverpool L3 5RF, UK}

\author[0000-0003-1546-6615]{Jesper~Sollerman}
\affiliation{\okc}

\author[0000-0001-6797-1889]{Steve~Schulze}
\affiliation{\okcp}

\author[0000-0001-6747-8509]{Yuhan~Yao}
\affiliation{\caltech}

\author[0000-0002-4223-103X]{Christoffer~Fremling}
\affiliation{\caltech}

\author[0000-0002-3653-5598]{Avishay~Gal-Yam}
\affiliation{\weizmann}

\author[0000-0002-9017-3567]{Anna~Y.~Q.~Ho}
\affiliation{\berkeley}
\affiliation{\millerfellow}

\author[0000-0002-6535-8500]{Yi~Yang}
\affiliation{\weizmann}
\affiliation{\berkeley}

\author[0000-0002-7252-3877]{Erik~C.~Kool}
\affiliation{\okc}

\author[0000-0002-7996-8780]{Ido~Irani}
\affiliation{\weizmann}

\author[0000-0003-1710-9339]{Lin~Yan}
\affiliation{\coo}

\author[0000-0003-3768-7515]{Igor Andreoni}
\affil{\caltech}

\author[0000-0003-1637-9679]{Dietrich Baade}
\affiliation{\eso}

\author[0000-0001-8018-5348]{Eric C. Bellm}
\affiliation{\dirac}

\author[0000-0001-5955-2502]{Thomas~G.~Brink}
\affiliation{\berkeley}

\author[0000-0002-1066-6098]{Ting-Wan~Chen}
\affiliation{\okc}

\author[0000-0001-7101-9831]{Aleksandar Cikota}
\affiliation{\esochile}
\affiliation{\lbnl}

\author[0000-0002-8262-2924]{Michael W. Coughlin}
\affiliation{\minnesota}

\author{Aishwarya Dahiwale}
\affiliation{\caltech}

\author[0000-0002-5884-7867]{Richard Dekany}
\affiliation{\coo}

\author[0000-0001-5060-8733]{Dmitry A. Duev}
\affiliation{\caltech}

\author[0000-0003-3460-0103]{Alexei V. Filippenko}
\affiliation{\berkeley}
\affiliation{\millerfellow}

\author[0000-0002-4338-6586]{Peter Hoeflich}
\affiliation{\fsu}

\author[0000-0002-5619-4938]{Mansi M. Kasliwal}
\affiliation{\caltech}

\author[0000-0001-5390-8563]{S.~R.~Kulkarni}
\affiliation{\caltech}

\author[0000-0001-9454-4639]{Ragnhild Lunnan}
\affiliation{\okc}

\author[0000-0002-8532-9395]{Frank J. Masci}
\affiliation{\ipac}

\author[0000-0003-0733-7215]{Justyn R. Maund}
\affiliation{\sheffield}

\author[0000-0002-7226-0659]{Michael S. Medford}
\affiliation{\berkeley}
\affiliation{\lbnl}

\author[0000-0002-0387-370X]{Reed Riddle}
\affiliation{\coo}

\author[0000-0002-6099-7565]{Philippe Rosnet}
\affiliation{\uca}

\author[0000-0003-4401-0430]{David L. Shupe}
\affiliation{\ipac}

\author[0000-0002-4667-6730]{Nora Linn Strotjohann}
\affiliation{\weizmann}

\author[0000-0003-0484-3331]{Anastasios Tzanidakis}
\affiliation{\caltech}

\author[0000-0002-2636-6508]{WeiKang Zheng}
\affiliation{\berkeley}

\begin{abstract}
We present observations of SN\,2021csp, the second example of a newly-identified type of supernova (Type Icn) hallmarked by strong, narrow, P~Cygni carbon features at early times.   
The SN appears as a fast and luminous blue transient at early times, reaching a peak absolute magnitude of $-20$ within 3 days due to strong interaction between fast SN ejecta ($v \approx 30000$\,km\,s$^{-1}$) and a massive, dense, fast-moving C/O wind shed by the WC-like progenitor months before explosion.  The narrow line features disappear from the spectrum 10--20 days after explosion and are replaced by a blue continuum dominated by broad Fe features, reminiscent of Type Ibn and IIn supernovae and indicative of weaker interaction with more extended H/He-poor material.  The transient then abruptly fades $\sim 60$\,days post-explosion when interaction ceases.  Deep limits at later phases suggest minimal heavy-element nucleosynthesis, a low ejecta mass, or both, and imply an origin distinct from that of classical Type Ic supernovae. 
We place SN\,2021csp in context with other fast-evolving interacting transients, and discuss various progenitor scenarios: an ultrastripped progenitor star, a pulsational pair-instability eruption, or a jet-driven fallback supernova from a Wolf-Rayet star.  The fallback scenario would naturally explain the similarity between these events and radio-loud fast transients, and suggests a picture in which most stars massive enough to undergo a WR phase collapse directly to black holes at the end of their lives.
\end{abstract}


\section{Introduction} 
\label{sec:intro}

Progenitor detections, hydrodynamic models, and basic rate calculations all suggest that most single stars born with initial masses of 8--20\,$M_\odot$ explode as red supergiants and produce Type IIP supernovae \citep[SNe;][]{Smartt2009}, but the fates of more-massive stars ($>$\,25\,$M_\odot$) remain an open question.  Such stars lose a significant fraction of their hydrogen (H) envelopes on the main sequence via line-driven winds even as single stars \citep[e.g.,][]{Vink2001}, and they are also more likely to undergo strong binary interaction \citep{Sana2012}.  In either case, a predicted consequence is that many such stars will be deficient in H by the time of core collapse.  Prior to explosion, these stars will appear as Wolf-Rayet (WR) stars; the explosion itself will then manifest as a supernova (SN) of spectroscopic Type IIb, Ib, or Ic (a stripped-envelope SN; for reviews, see \citealt{Filippenko1997} and \citealt{GalYam2017}).

This straightforward picture faces a number of challenges, however.  First, hydrodynamic models suggest that the masses ejected by typical SNe~Ib/c are only a few $M_\odot$, much lower than predicted for exploding WR stars \citep[e.g.,][]{Dessart2012}.
Second, no WR star has yet been identified at the site of a SN in pre-explosion imaging: the handful of reported SN~Ib/c progenitor candidates are too optically luminous to be WR stars \citep{Cao+2013,Eldridge2016,vanDyk2018,Kilpatrick+2018,Kilpatrick+2021}, and upper limits on the remainder are in marginal tension with the luminosity distribution of the Galactic WR population (\citealt{Eldridge2013}, although cf.\ \citealt{Maund2018} and \citealt{Sander2019}).  Third, SNe~Ib/c are too abundant (by a factor of $\sim 2$) to originate solely from the WR population \citep{Smith2011}.

For these reasons, binary evolution involving pairs of lower-mass stars undergoing a common-envelope phase has increasingly been seen as the most likely pathway for explaining most of the SN~Ib/c population.  If so, the final outcome of stellar evolution for more-massive stars ($\gtrsim 25\,M_\odot$) remains unclear.  One possibility is that very massive stars do not explode at all, and instead collapse directly to black holes with minimal emission of electromagnetic radiation \citep{OConnor2011,Sukhbold2014,Smartt2015,Zapartas2021}.  This remains controversial.  Some very massive stars probably explode while still in possession of their H envelope to produce SNe~IIn \citep{GalYam2007,GalYam2009,Smith2011,Mauerhan2013,Smith2014}, although this does not resolve the question of the fates of those massive stars that do undergo a WR phase.  Some \emph{atypical} SNe~Ib/c do appear to be consistent with massive WR progenitors: specifically, about 25\% of broad-lined SNe~Ic (Ic-BL) show ejecta masses consistent with explosions of very massive stars \citep{Taddia2019}, and the progenitors of superluminous SNe are also likely to be quite massive \citep{GalYam2009b,Nicholl2015,Jerkstrand2017,Blanchard2020}. 

Another rare stripped-envelope SN subtype that has been suggested to be related to very massive stars is the class of Type Ibn SNe.
The velocities inferred from the widths of the hallmark narrow helium (He) lines of these systems --- attributed to dense circumstellar matter (CSM) surrounding the progenitor star --- are comparable to those seen in Local Group WR stars, suggesting that WR stars may indeed be their progenitors \citep{Foley2007,Pastorello2008}.  However, the pre-explosion mass-loss rates inferred from observations of SNe~Ibn are much higher than those seen in normal WR winds, implying that any WR progenitor must enter a short evolutionary phase of greatly enhanced mass loss prior to the explosion.

The list of stripped-envelope SN subtypes continues to expand.  \cite{GalYam2021} recently presented a detailed observational study of SN\,2019hgp, a fast and luminous transient with no known literature precedent.  Early-time spectra of this event are dominated by narrow lines with profiles similar to those seen in SNe~Ibn but originating from carbon (C), oxygen (O), and other alpha elements rather than He, defining a new class of ``Type Icn'' SNe that previously was only theoretical \citep{Smith2017,Woosley2017}.  In their analysis of this object, \cite{GalYam2021} point out that the distinction between SNe~Ibn and SNe~Icn closely mirrors that of the WR spectroscopic subtypes (He/N-rich WN versus He-poor, C-rich WC stars).  On this basis, they postulate that SNe~Ibn/Icn represent the true outcomes of the explosions of WR stars.

\cite{GalYam2021} also note that the properties of SN\,2019hgp (fast-rising, hot, and luminous) show some resemblance to the population of rare, fast-evolving transients identified in photometric surveys (\citealt{Drout2014,Arcavi2016,Tanaka2016,Pursiainen2018,Inserra2019,Ho2021}), sometimes referred to as fast blue optical transients (FBOTs) or rapidly evolving transients (RETs), indicating a possible link with this previously poorly-explored group of objects.  However, SN\,2019hgp would not itself have been classified as an FBOT/RET by the criteria typically employed in earlier works.

In this paper, we present observations of the second SN~Icn  to be discovered, SN\,2021csp.
The properties of this object are qualitatively similar to those of SN\,2019hgp but even more extreme. SN\,2021csp is faster and more luminous, and a far more extensive observational campaign was possible.   These observations strengthen the basic model presented by \cite{GalYam2021} but also allow us to further extend it, with important implications for the fates of very massive stars of all types.
Indeed, we argue that the distinction between SNe~Ibn/Icn and ``normal'' SNe~Ib/Ic may involve not only the mass and evolutionary history of the progenitor, but also the nature of the underlying explosion and the type of compact remnant that is left behind.

The paper is organized as follows.  Section~\ref{sec:observations} presents the discovery of SN\,2021csp and our extensive observational campaign. We perform a more detailed analysis of the light curve, spectra, and host galaxy in \S \ref{sec:analysis} to infer some basic properties of the explosion and pre-explosion system.  In \S \ref{sec:discussion} we discuss the results of the analysis in the context of the physical nature of the progenitor, its evolutionary state prior to explosion, and the nature of the explosion itself. Section~\ref{sec:interpretation} discusses the implications of these results for progenitor models, and \S \ref{sec:conclusions} summarizes our conclusions.   We use a standard cosmological model with H$_0 = 70$\,km\,s$^{-1}$\,Mpc$^{-1}$, $\Omega_\Lambda = 0.7$, and $\Omega_M = 0.3$, corresponding to a distance modulus of $\mu = 37.91$\ mag at the redshift of SN\,2021csp ($z=0.084$; \S \ref{sec:sprat}).
UT dates are used throughout, and times of observations are referenced to an estimated explosion date of MJD 59254.5 (\S \ref{sec:exptime}) and expressed in the rest frame.  Apparent magnitudes are reported in the text in the AB system \citep{Oke1983} without an extinction correction, but for analysis and in our figures we correct for Galactic extinction assuming a reddening of $E(B-V) = 0.027$\ mag \citep{Schlafly2011}.

\section{Observations}
\label{sec:observations}

\subsection{Palomar 48-inch Discovery and Photometry}
\label{sec:p48}

The Zwicky Transient Facility (ZTF; \citealt{Bellm2019,Graham2019}) is a combined public and private time-domain optical sky survey, using a 47 deg$^{2}$ field-of-view camera \citep{Dekany2020}
on the refurbished Samuel Oschin 48-inch Schmidt telescope (P48) 
at Palomar Observatory.  The ZTF observing and alert system are described in previous works \citep{Masci2019,Patterson2019,Mahabal2019,Duev2019}.  

SN\,2021csp (internally designated ZTF21aakilyd) was first detected in an $i$-band image obtained on 2021-02-11 (MJD 59256.4766) as part of the ZTF high-cadence survey
\citep{Bellm2019surveys} and confirmed with a second observation in the $g$ band the same night.  The last nondetection was two  days prior.  It was identified as a candidate of interest the following morning during daily scanning of our custom alert filter \citep{Ho2020koala,Perley2021}, owing to the fast rise ($> 2.5$\ mag in two days), blue colors ($g-i = -1$\ mag), and coincidence with an extended object (the host galaxy), motivating a substantial follow-up campaign (\S \ref{sec:imaging}--\ref{sec:multiwavelength}).

We used the IPAC forced-photometry pipeline \citep{Masci2019} to obtain final P48 photometry and pre-explosion upper limits, reported in Table~\ref{tab:photometry}.  A long sequence of ultra-high-cadence imaging from 2021-02-18 (100 consecutive 30\,s exposures, followed by another 26 consecutive 30\,s exposures) has been averaged together into two measurements.

\begin{deluxetable}{lcccccc}
\tablewidth{0pt}
\tabletypesize{\footnotesize}
\tablecaption{Photometry of SN\,2021csp}
\tablehead{
\colhead{Telescope} & 
\colhead{MJD} & 
\colhead{Filter} & 
\colhead{AB Mag} & 
\colhead{unc.} \\
\colhead{} & 
\colhead{(days)} &
\colhead{} & 
\colhead{} & 
\colhead{}
}
\startdata
P48   & 59250.4258 & {\it r   } & $>$21.09 &  \\
P48   & 59250.4648 & {\it g   } & $>$21.27 &  \\
P48   & 59252.4141 & {\it i   } & $>$20.72 &  \\
P48   & 59252.5195 & {\it r   } & $>$21.75 &  \\
P48   & 59254.4219 & {\it r   } & $>$20.79 &  \\
P48   & 59254.5273 & {\it g   } & $>$21.50 &  \\
P48   & 59256.4766 & {\it i   } &  19.05 & 0.06 \\
P48   & 59256.5078 & {\it g   } &  18.11 & 0.03 \\
LT    & 59257.1992 & {\it g   } &  17.92 & 0.03 \\
LT    & 59257.1992 & {\it r   } &  18.25 & 0.03 \\
LT    & 59257.2031 & {\it u   } &  17.53 & 0.03 \\
LT    & 59257.2031 & {\it i   } &  18.59 & 0.03 \\
LT    & 59257.2031 & {\it z   } &  18.84 & 0.03 \\
UVOT  & 59257.9570 & {\it UVW1} &  17.43 & 0.06 \\
UVOT  & 59257.9688 & {\it UVW2} &  17.36 & 0.07 \\
UVOT  & 59257.9766 & {\it UVM2} &  17.27 & 0.05 \\
LT    & 59258.1367 & {\it g   } &  17.82 & 0.03 \\
LT    & 59258.1406 & {\it i   } &  18.43 & 0.03 \\
LT    & 59258.1406 & {\it u   } &  17.52 & 0.03 \\
LT    & 59258.1406 & {\it r   } &  18.09 & 0.03 \\
LT    & 59258.1445 & {\it z   } &  18.62 & 0.05 \\
\enddata
\label{tab:photometry}
\tablecomments{Magnitudes are not corrected for Galactic extinction.  Only the first few entries are provided here: a complete machine-readable table is available online.}
\end{deluxetable}

We also conducted a more extensive search of the pre-explosion P48 data (extending back to the start of the ZTF survey in 2018, 3\,yr prior to the explosion time) to search for precursor outbursts, following the procedure described by \cite{Strotjohann2021}.  No significant detections prior to the explosion date were found, to typical (median) limits of $\sim 20.5$\ mag ($-17.5$ absolute magnitude) in 1\ day bins or to $\sim 22$\ mag ($-16$ absolute magnitude) in bins up to 90\ days in width. 
Previously-observed outbursts prior to interacting SNe~Ibn and SNe~IIn have ranged in luminosity between $-12.5$ and $-17$ mag \citep{Strotjohann2021}, with only five previously-reported cases of outbursts more luminous than $-16$ mag among 143 SNe of these types searched using this procedure.
These limits therefore rule out only the most luminous potential outbursts.

\subsection{Imaging}
\label{sec:imaging}

\subsubsection{Liverpool Telescope}
\label{sec:ioo}
\begin{figure}
    \centering
    \includegraphics[width=0.44\textwidth]{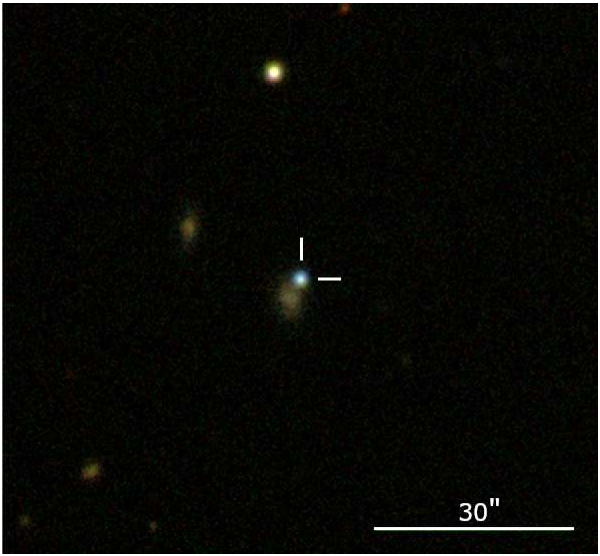}
    \caption{A false-color $gri$ image of the field from 2021-02-11 taken with IO:O on the LT. North is up and east to the left.  SN\,2021csp is seen as a blue source in an outer spiral arm of its host galaxy, northwest of the nucleus.
    }
    \label{fig:image}
\end{figure}

We obtained $ugriz$ imaging using the Infrared/Optical Imager (IO:O) on the 2\,m robotic Liverpool Telescope (LT; \citealt{Steele2004}) starting from the first night following the discovery and continuing until the object faded below detection (55\ days later).
Data were processed by the IO:O automatic pipeline and obtained in reduced form from the LT archive.  We subtracted reference imaging from Pan-STARRS ($griz$ bands) or from the Sloan Digital Sky Survey (SDSS; $u$ band) using a custom IDL subtraction pipeline, and performed seeing-matched aperture photometry.  A color image of the field is shown in Figure~\ref{fig:image}.

\subsubsection{Palomar 60-inch Telescope}
\label{sec:p60rc}
We obtained additional $ugri$ photometry using the Rainbow Camera of the Spectral Energy Distribution Machine (SEDM; \citealt{Blagorodnova2018}) on the robotic Palomar 60-inch telescope (P60; \citealt{Cenko2006}).  Image subtraction and photometry was performed using {\tt FPipe} \citep{Fremling2016}.

\subsubsection{Swift Ultraviolet/Optical Telescope}
\label{sec:uvot}
We observed the field of SN\,2021csp with the Ultraviolet/Optical Telescope (UVOT; \citealt{Roming2005a}) on board the {\it Neil Gehrels Swift Observatory} \citep{Gehrels2004a} beginning 2021-02-12 and continuing until the flux from the transient faded below detectability a month later.  An additional set of observations between 2021-03-31 and 2021-04-21 were acquired to constrain the host-galaxy flux.  The brightness in the UVOT filters was measured with UVOT-specific tools in the HEAsoft\footnote{\url{https://heasarc.gsfc.nasa.gov/docs/software/heasoft/}} version 6.26.1 \citep{Blackburn1995}. Source counts were extracted from the images using a circular aperture of radius $3''$. The background was estimated over a significantly larger area close to the SN position. The count rates were obtained from the images using the \swift\ tool {\tt uvotsource}. They were converted to AB magnitudes using the UVOT photometric zero points of \cite{Breeveld2011a} and the UVOT calibration files from September 2020. 
To remove the host emission from the transient light curves, we used templates formed from our final observations in April and from archival UVOT observations of the field from 2012.  We measured the host contribution using the same source and background apertures, and subtracted this contribution from the transient flux measurements.

\subsubsection{Nordic Optical Telescope}
\label{sec:notim}

We obtained five epochs of imaging with the Alhambra Faint Object Spectrograph and Camera (ALFOSC) on the 2.56\,m Nordic Optical Telescope (NOT).  Observations were obtained on 2021-04-03, 2021-04-18, 2021-04-20, 2021-05-07, and 2021-07-01.  For the first two epochs, $gri$ observations were obtained; for the last three epochs only deep $r$-band observations were acquired.  All observations were taken under clear skies and subarcsecond seeing except the data from 2021-04-18 which were affected by thin clouds and relatively poor seeing ($\sim1\farcs3$). 
Data were reduced with the Python package PyNOT\footnote{\href{https://github.com/jkrogager/PyNOT}{https://github.com/jkrogager/PyNOT}} (v0.9.7).

\begin{figure}
    \centering
    \includegraphics[width=0.47\textwidth]{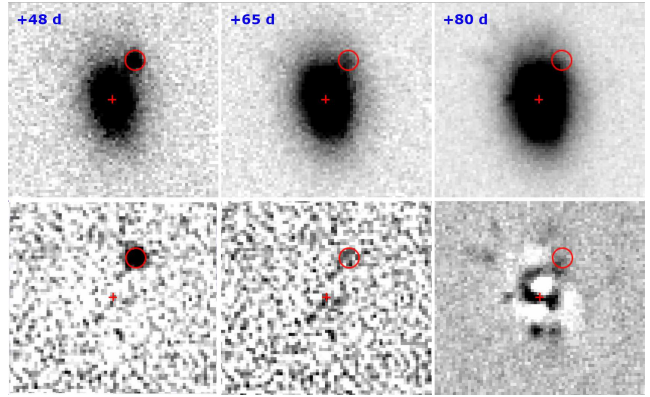}
    \caption{NOT imaging of SN\,2021csp during the rapid late-phase light-curve decline.  The top row shows the images without image subtraction; the bottom row displays images after subtraction of the host galaxy.  The center of the host galaxy is marked with a cross and the position of the SN with a circle of 0$\farcs$7 radius in all images.  Pan-STARRS imaging has been used to subtract the images at +48 and +65\ days, although the subtraction at +65\ days is limited by the depth of the reference.  GALFIT has been used to subtract an axisymmetric model of the host galaxy to obtain the image at lower right.  No source is detected at the SN location in this image.
    }
    \label{fig:image3}
\end{figure}

For the three sets of observations taken in April, we employ Pan-STARRS templates for subtraction using the same methods used for the LT photometry.  
By the time of the observation in May, the transient had faded to a very faint level and this method was no longer sufficient.  While a secure limit of $r > 23.66$\ mag can be obtained from the Pan-STARRS subtraction, this is limited entirely by the depth of the reference 
(the true 3$\sigma$ limiting magnitude of this image, measured away from the galaxy, is $r \approx 26.2$\ mag).
Instead, we employ the software utility GALFIT \citep{Peng2002,Peng2010} to model the disk of the galaxy as a S\'ersic profile (convolved with the point-spread function) and remove it from our images.  The model provides only incomplete removal of the inner galaxy light, and the inner spiral pattern and \ion{H}{2} regions are visible as residuals in the subtracted image.  However, the immediate vicinity around the location of the transient does not show any major residuals 
(Fig.~\ref{fig:image3}), including any evidence of light from the transient. 
Forced photometry at the transient location gives $r=25.4 \pm 0.15$\ mag, although the flux is probably dominated by light from an unsubtracted \ion{H}{2} region centered just outside the aperture.  As a conservative upper limit, we report $r>24.8$\ mag in our photometry table (corresponding to 5$\sigma$ above the forced-photometry value in flux units).

The observation from July is not as deep as the one obtained in May, and so is not individually constraining.  To confirm the accuracy of our GALFIT subtraction, we carried out image subtraction between the May and July observations and obtained an upper limit (difference magnitude) of $r>25.1$\ mag (3$\sigma$).  However, since we cannot rule out the possibility that a small amount of flux is present in the July observation, we will generally use the more conservative GALFIT-based approach.

\begin{figure*}
    \centering
    \includegraphics[width=0.98\textwidth]{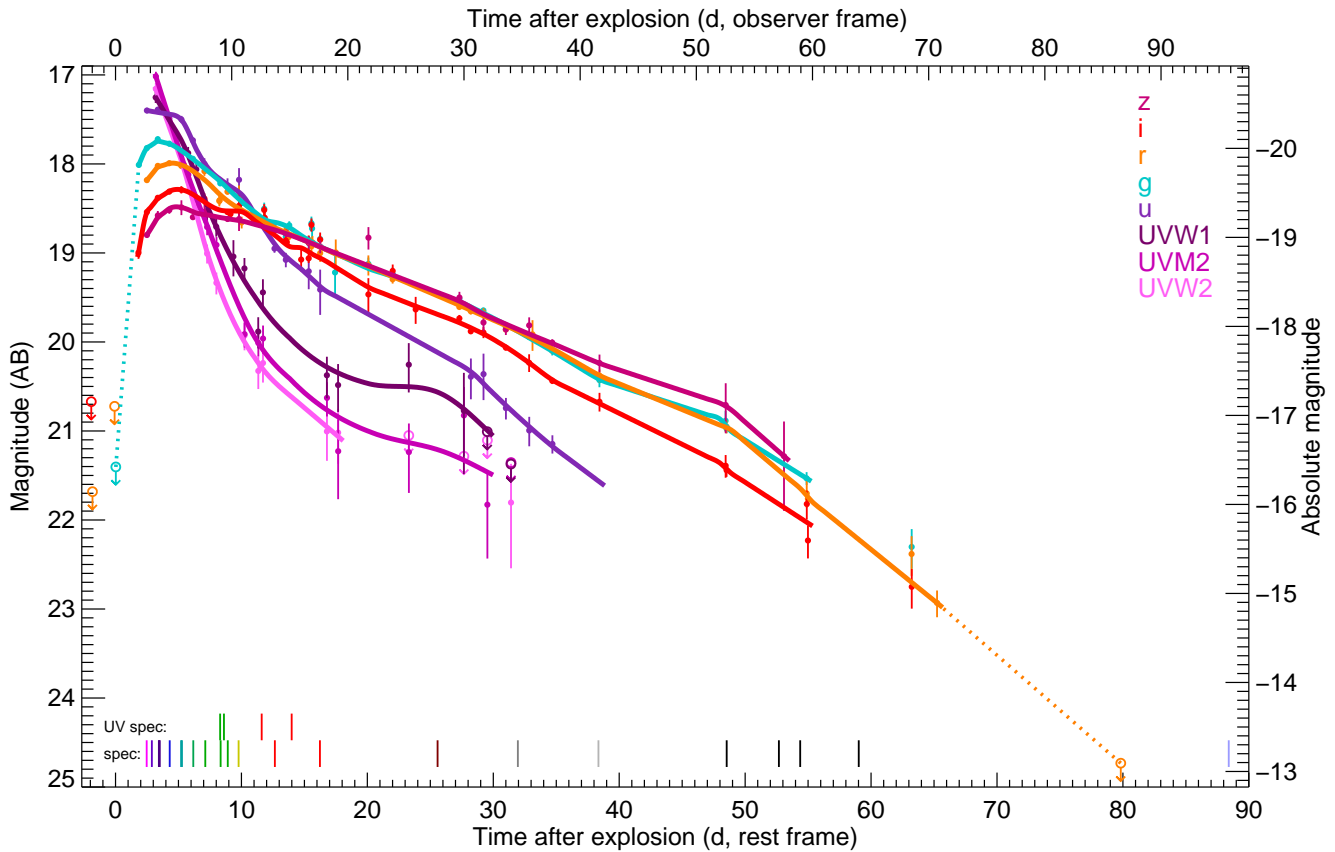}
    \caption{The ultraviolet/optical/near-infrared light curves of SN\,2021csp.  The transient reached a peak absolute magnitude of $M_g \approx -20.1$\ mag (and a bolometric luminosity $L_{\rm bol} > 10^{44}$\,erg\,s$^{-1}$) within 4\ days of explosion and then rapidly faded, qualifying it as one of the most nearby examples of an FBOT.  Interpolation curves for each filter band are estimated using a combination of local regression and spline fitting.
    Dotted lines connect the most constraining upper limit with the first detection in the same band, and the last detection with the first subsequent deep upper limit.
    Bars at the bottom indicate observation times of spectroscopy.
    }
    \label{fig:lightcurve}
\end{figure*}

A light curve showing the P48, LT, P60, NOT, and UVOT photometry of SN\,2021csp is shown in Figure~\ref{fig:lightcurve}.

\subsection{Spectroscopy}
\label{sec:spectroscopy}

\begin{figure}
    \centering
    \includegraphics[width=0.45\textwidth]{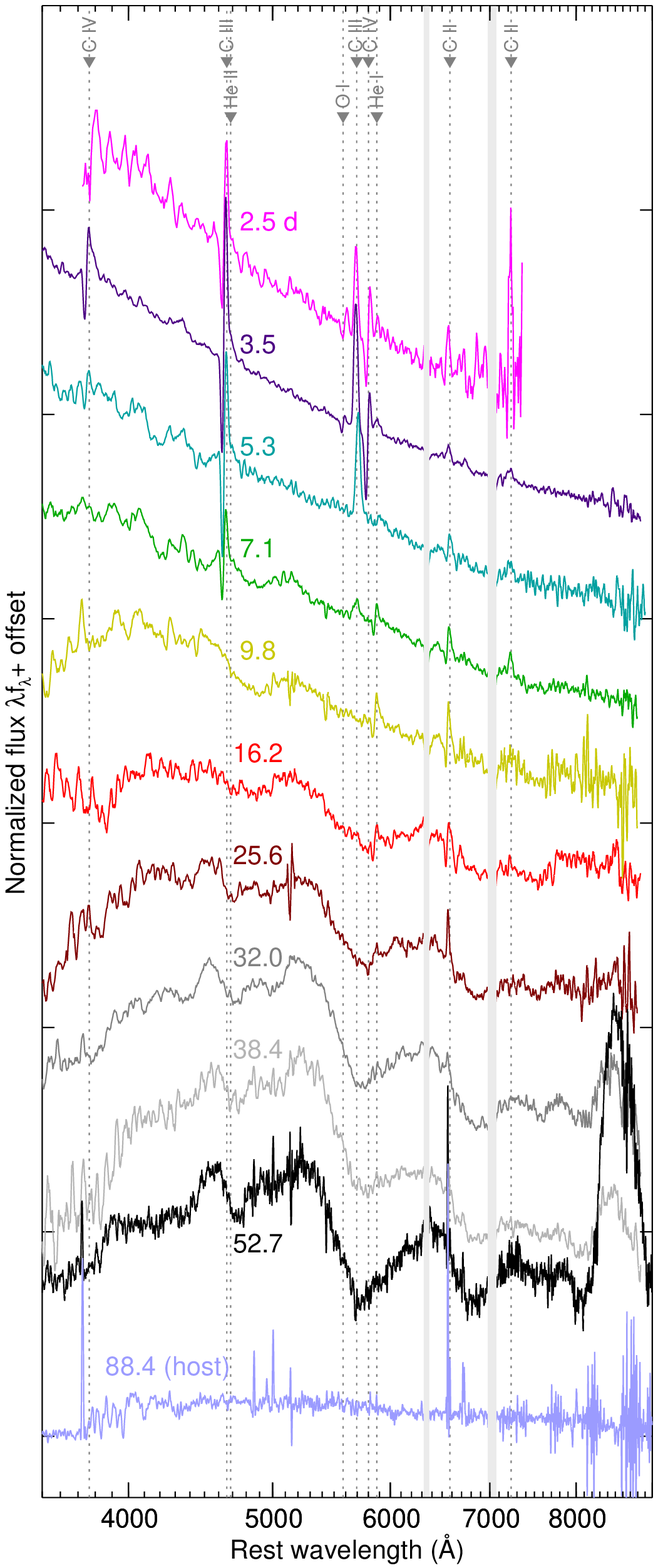}
    \caption{Spectral evolution of SN\,2021csp from selected observations (see Table \ref{tab:spec_tab} for the full observing log).  Labels indicate the rest-frame time since the assumed explosion date and prominent CSM features are labeled.  Early-phase spectra are dominated by narrow P~Cygni features of ionized C, which vanish 8--12\ days post-explosion (rest frame).  Broad lines are dominant at 15--53\ days.  The final spectrum at 88.4\ days shows only host-galaxy flux and is normalized on the same scale as the 52.7\ day spectrum.
    }
    \label{fig:optspec}
\end{figure}

We obtained an extensive series of optical spectra beginning prior to the peak of the SN and extending until 80\ days post-explosion in the rest frame.  A log of all spectroscopic observations, 25 epochs in total, is provided in Table~\ref{tab:spec_tab} and the spectra will be uploaded to WISEREP\footnote{https://www.wiserep.org} \citep{Yaron2012}.  Details of the observations are provided below.  In addition, we use our $g$- and $r$-band light curves to perform an absolute calibration and color-correction on each spectrum.  We calculate synthetic magnitudes of each (flux-calibrated, pre-corrected) spectrum in both filters and apply a rescaling (to match the absolute fluxes) followed by a power-law correction (to match the colors).  At late times ($>50$\ days) we apply only the absolute scaling with no color correction.  A time series including many of the spectral observations is displayed in Figure~\ref{fig:optspec}.

\begin{deluxetable*}{lccccccc}
\tablewidth{0pt}
\tabletypesize{\footnotesize}
\tablecaption{Log of spectroscopic observations of SN\,2021csp}
\tablehead{
\colhead{Observation date} & 
\colhead{MJD} &
\colhead{Phase} &
\colhead{Facility} &
\colhead{Exp. time} &
\colhead{Grism/Grating} &
\colhead{Slit width} &
\colhead{Range} \\
\colhead{(UTC)} & 
\colhead{(days)} &
\colhead{(days)} &
\colhead{} &
\colhead{(s)} &
\colhead{} &
\colhead{(arcsec)} &
\colhead{(\AA)} 
}
\startdata
2021 Feb 12	04:23:54 & 59257.183 & 2.475 & LT/SPRAT	   & $2\times600$ & Blue		& 1.8 & 4020--7994 \\
2021 Feb 12	15:07:38 & 59257.630 & 2.888 & Gemini/GMOS & $2\times900$ & B600		& 1.0 & 3641--6878 \\
2021 Feb 13 05:36:01 & 59258.233 & 3.444 & LT/SPRAT    & $2\times600$ & Blue        & 1.8 & 4000--8000 \\
2021 Feb 13	06:18:33 & 59258.263 & 3.471 & NOT/ALFOSC  & 1800		  &	Grism\#4	& 1.0 &	3852--9681 \\ 
2021 Feb 13 07:18:40 & 59258.305 & 3.510 & VLT/FORS2   & $8\times750$ & 300V          & 1.0 & 4400--9200 \\
2021 Feb 14 03:54:36 & 59259.163 & 4.302 & LT/SPRAT    & $2\times600$ & Blue        & 1.8 & 4000--8000 \\
2021 Feb 15 03:30:05 & 59260.146 & 5.208 & LT/SPRAT    & $2\times600$ & Blue        & 1.8 & 4000--8000 \\
2021 Feb 15	05:17:55 & 59260.221 & 5.277 & NOT/ALFOSC  & $2\times900$ & Grism\#4	& 1.3 &	3501--9635 \\ 
2021 Feb 16 04:40:52 & 59261.195 & 6.176 & LT/SPRAT    & $2\times600$ & Blue        & 1.8 & 4000--8000 \\
2021 Feb 17	05:24:30 & 59262.225 & 7.127 & NOT/ALFOSC  & $2\times1800$& Grism\#4	& 1.3 &	3504--9635 \\ 
2021 Feb 18 12:09:43 & 59263.507 & 8.309 & HST/STIS    & 2100 & G230L & 0.2 & 1570--3180 \\        
2021 Feb 18	13:12:00 & 59263.550 & 8.349 & Lick/Kast  & $1\times3660$ & 600/4310   & 2.0 &	3632--10340	\\
                     &           &       &             & $3\times1200$ & 300/7500   &     &             \\
2021 Feb 18 20:05:29 & 59263.837 & 8.614 & HST/COS     & 4243 & G140L & 3.0 & 1230--2050 \\ 
2021 Feb 19 03:57:39 & 59264.165 & 8.916 & LT/SPRAT    & $2\times600$ & Blue        & 1.8 & 4000--8000 \\
2021 Feb 20	02:16:36 & 59265.095 & 9.774 & NOT/ALFOSC  & $2\times1800$&	Grism\#4	& 1.3 &	3501--9631 \\ 
2021 Feb 22 01:58:28 & 59267.082 &11.607 & HST/STIS    & 2030 & G230L & 0.2 & 1570--3180 \\
2021 Feb 23 05:15:11 & 59268.219 &12.656 & LT/SPRAT    & $2\times600$ & Blue        & 1.8 & 4000--8000 \\
2021 Feb 24 15:58:17 & 59269.665 &13.990 & HST/COS     & 4003 & G140L & 3.0 & 1230--2050 \\
2021 Feb 27	02:19:35 & 59272.097 &16.233 & NOT/ALFOSC  & $2\times1800$&	Grism\#4	& 1.0 &	3753--9683 \\ 
2021 Mar 09	05:19:36 & 59282.222 &25.574 & NOT/ALFOSC  & $2\times1800$&	Grism\#4	& 1.3 &	3752--9620 \\ 
2021 Mar 16	03:18:45 & 59289.138 &31.954 & NOT/ALFOSC  & $3\times1500$&	Grism\#4	& 1.0 &	3701--9683 \\ 
2021 Mar 23	01:48:34 & 59296.075 &38.354 & NOT/ALFOSC  & $3\times1500$&	Grism\#4	& 1.0 &	4001--9685 \\ 
2021 Apr 03 02:42:56 & 59307.113 &48.536 & NOT/ALFOSC  & $3\times1500$&	Grism\#4	& 1.0 &	4003--9677 \\ 
2021 Apr 07 14:32:41 & 59311.606 &52.681 & Keck/LRIS   & 850+750      & B600/4000   & 1.0 & 3134--10284\\
                     &           &       &             & 750+670      & R400/8500   &     &            \\
2021 Apr 09 10:28:48 & 59313.437 &54.370 & Palomar/DBSP& $1200$       & B600        & 1.5 & 3400--10000 \\
                     &           &       &             & $1200$       & R316        &     &            \\
2021 Apr 14 11:32:20 & 59318.481 &59.023 & Keck/LRIS   & $450$        & B400/3400   & 1.0 & 3000--10306\\
                     &           &       &             & $450$        & R400/8500   &     &            \\
2021 May 10 10:56:35 & 59344.456 &82.985 & Keck/LRIS   & $3\times900$ & B400/3400   & 1.0 & 3000--10306\\ 
                     &           &       &             & $3\times903$ & R400/8500   &     &            \\
2021 May 16 07:52:14 & 59350.328 &88.402 & Keck/LRIS   & $3\times900$ & B400/3400   & 1.0 & 3000--10306\\
                     &           &       &             & $3\times903$ & R400/8500   &     &            \\
\enddata
\label{tab:spec_tab}
\tablecomments{The phase is calculated with respect to MJD 59254.5 (the estimated explosion date) and is given in the rest frame.}
\end{deluxetable*}

\subsubsection{Liverpool Telescope}
\label{sec:sprat}
We obtained seven sets of spectra (each 2$\times$600\,s) spanning the first two weeks after explosion using the Spectrograph for the Rapid Acquisition of Transients (SPRAT; \citealt{Piascik2014}).  We use the default reduction and extraction provided by the SPRAT pipeline \citep{Barnsley2012}.   The first LT spectrum immediately established the redshift and unusual nature of this transient on the basis of the detection of several strong C features at a common redshift of $z=0.084$ \citep{Perley2021_astronote}.  The only known previous event sharing these features is SN\,2019hgp \citep{GalYam2021}, although a third event (SN\,2021ckj, also discovered using ZTF data) was reported a few days later \citep{Pastorello2021_AN}.  The strong similarities between these events (and their H/He-free narrow-line spectra at maximum light) motivated us to propose these three events as the prototypical members of the new class of Type Icn SNe \citep{GalYam2021_AN}.

\subsubsection{Gemini-North}
\label{sec:gmos}
One spectrum was obtained on 2021-02-12 with the Gemini Multi-Object Spectrograph (GMOS; \citealt{Hook2004}) mounted on the Gemini North 8\,m telescope at the Gemini Observatory on Maunakea, Hawaii. Two 900\,s exposures were obtained with the B600 grating. 
The GMOS data were reduced using {\tt PypeIt} \citep{Prochaska2020}.

\subsubsection{Nordic Optical Telescope}
\label{sec:notspec}
We obtained nine separate epochs of spectroscopy with the ALFOSC on the NOT spanning from 2021-02-13 until 2021-04-03 (Table~\ref{tab:spec_tab}).  Observations were taken using Grism \#4, providing wavelength coverage over most of the optical spectral range (typically 3700--9600\,\AA).  The slit was aligned with the parallactic angle \citep{Filippenko1982}, except in the last three observations when it also included the host, and an atmospheric dispersion corrector was used. 
Reduction and calibration were performed using {\tt PypeIt}.

\subsubsection{Very Large Telescope}
\label{sec:fors2}
Spectropolarimetry of SN\,2021csp was conducted with the FOcal Reducer and 
low dispersion Spectrograph (FORS2; \citealt{Appenzeller1992})
on Unit Telescope\,1 (UT1, Antu) of the ESO Very Large Telescope (VLT). The observations were carried out in the Polarimetric Multi-Object Spectroscopy (PMOS) mode on 2021-02-13. 
Two sets of data were obtained, each consisting of four 750\,s exposures with the retarder oriented at angles of 0, 22.5, 45 and 67.5 degrees.  The 300V grism and the $1''$ wide slit were used, yielding a spectral resolving power of $R \approx 440$ at the central wavelength of 5849\,\AA.  The GG435 order-separation filter was used to minimize second-order contamination at redder wavelengths.  This configuration provided a wavelength coverage of about 4400--9200\,\AA\ in the observer frame.  The total Stokes $I$ spectrum was flux calibrated based on the observation of a spectrophotometric standard star with the polarimetric optics in place, and the retarder plate set to $0^\circ$.

The data were bias subtracted and flat-field corrected. For each individual 
exposure, the ordinary (o) and extraordinary (e) beams were extracted and 
wavelength-calibrated separately following standard procedures within 
IRAF \citep{Tody1986}.  After removing the instrumental polarization of FORS2 
\citep{Cikota2017}, we derived the Stokes parameters, the 
bias-corrected polarization, and the associated uncertainties using our own 
routines, following the procedures of \cite{Patat2006}, \citet{Maund2007}, \citet{Simmons1985}, and \citet{Wang1997}. A detailed 
description of the reduction of FORS spectropolarimetry is given by 
\citet[][their Appendix A]{Yang2020}.

\begin{figure}
    \centering
    \includegraphics[width=0.47\textwidth]{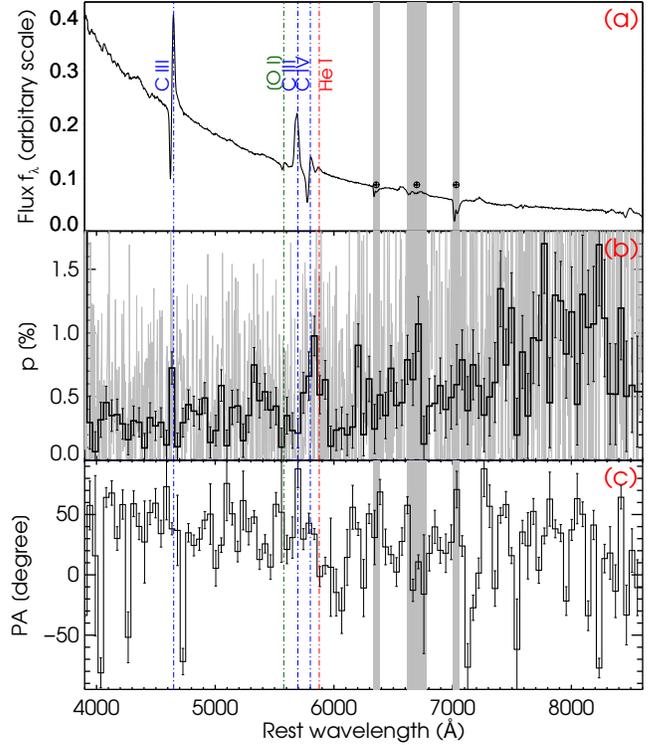}
    \caption{Spectropolarimetry of SN\,2021csp at $\sim 3.5$\ days (rest-frame). The three 
    panels (from top to bottom) show (a) the scaled total-flux spectrum 
    with \ion{C}{3}, \ion{C}{4}, [\ion{O}{1}], and \ion{He}{1} lines 
    labeled at zero velocity relative to the SN; 
    (b) the polarization spectrum ($p$); 
    and (c) the polarization position angle (PA). Vertical gray-shaded 
    regions indicate the major telluric features.  The data have been rebinned to 
    50\,\AA\ to improve the S/N. 
    }
    \label{fig:specpol}
\end{figure}

The Stokes parameters computed for each set of four exposures are 
consistent with each other. We further combined the two beams for o-ray 
and e-ray at each retarder angle and derived the Stokes parameters. The 
intensity-normalized Stokes parameters ($I$, $Q$, $U$) are in 50\,\AA\ wide bins ($\sim 22$ pixels) to further increase the signal-to-noise 
ratio (S/N). The results are presented in Figure~\ref{fig:specpol}. 

\subsubsection{Lick 3\,m Telescope}
\label{sec:kast}

A single optical spectrum of SN\,2021csp was obtained with the Kast double spectrograph \citep{Miller1993} mounted on the 3\,m Shane telescope at Lick Observatory.  The spectrum was taken at or near the parallactic angle \citep{Filippenko1982} to minimize slit losses caused by atmospheric dispersion. Data were reduced following standard techniques for CCD processing and spectrum extraction \citep{Silverman2012} utilizing IRAF routines and custom Python and IDL codes\footnote{https://github.com/ishivvers/TheKastShiv}.  Low-order polynomial fits to arc-lamp spectra were used to calibrate the wavelength scale, and small adjustments derived from night-sky lines in the target frames were applied.  Observations of appropriate spectrophotometric standard stars were used to flux calibrate the spectra.

\subsubsection{Hubble Space Telescope}

We obtained two sets of observations of SN\,2021csp with the \emph{Hubble Space Telescope} (\emph{HST}), using both the Cosmic Origins Spectrograph (COS; \citealt{Green2012}) and the Space Telescope Imaging Spectrograph (STIS; \citealt{Woodgate1998})\footnote{program ID GO\#16212 (PI Perley)}.  The COS observations employed the G140L grating and the STIS observations used the G230L grating.  The first set of observations was taken at 8.31 and 8.61 rest-frame days after our assumed explosion time (for STIS and COS, respectively); the second set was taken at 11.61\ days (STIS) and 13.99\ days (COS). 

\begin{figure*}
    \centering
    \includegraphics[width=0.96\textwidth]{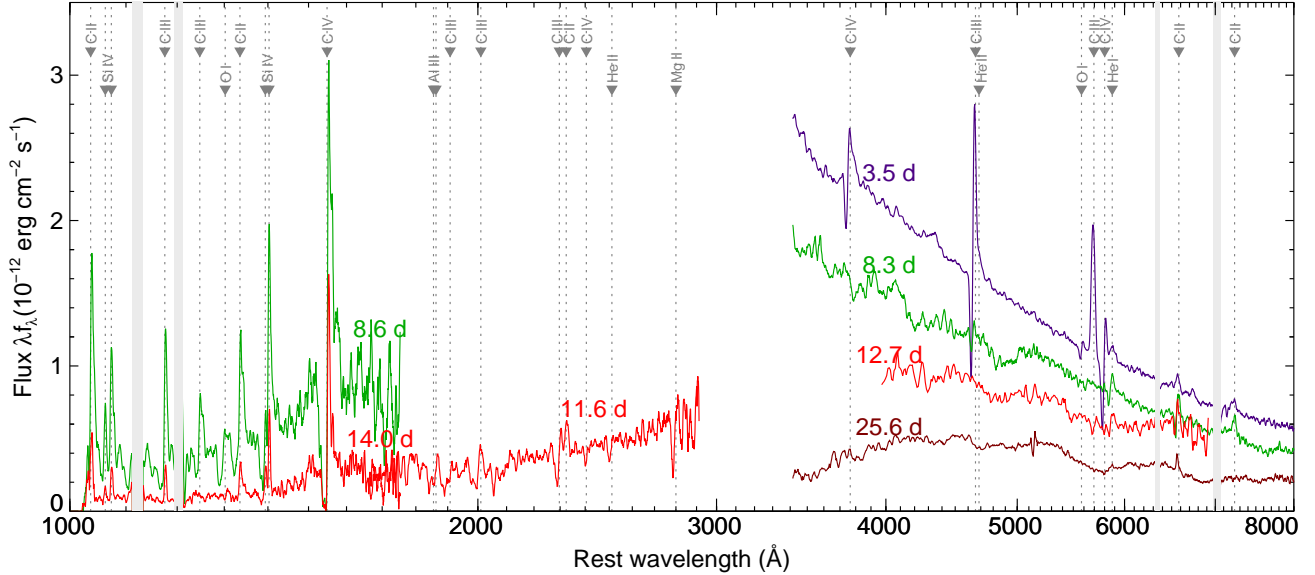}
    \caption{Combined UV-optical spectral series showing the relative strengths of the narrow emission features at various approximately coeval epochs (all times are rest-frame days from the assumed explosion time).  Identified transitions are marked with dotted lines, and regions of strong telluric absorption or geocoronal contamination are marked with gray bands.
    }
    \label{fig:uvoptspec}
\end{figure*}

We use the pipeline reductions from the \emph{HST} archive.  The first STIS spectrum has S/N about a factor of 10 lower than expected, likely due to a guiding problem.  This problem is not seen in the second STIS exposure or with COS.  The UV spectra are shown alongside optical spectra obtained at similar times in Figure~\ref{fig:uvoptspec}.

\subsubsection{Palomar 200-inch Telescope}

One spectrum of SN\,2021csp was acquired with the Double Beam Spectrograph (DBSP; \citealt{Oke1982}) on the 5\,m Hale telescope at Palomar Observatory (P200). Observations were taken on 2021-04-09, using the 600/4000 grating on the blue side and the 316/7150 grating on the red side.  Data were reduced using the DBSP-DRP fully-automated pipeline\footnote{\url{https://github.com/finagle29/dbsp_drp}}.

\subsubsection{Keck Observatory}

Spectra of SN\,2021csp were acquired on four separate occasions with the Low Resolution Imaging Spectrometer (LRIS; \citealt{Oke1995}) on the Keck~I 10\,m telescope.  The first observation was obtained on 2021-04-07 using the B600/4000 blue-side grism and the R400/8500 red-side grating; the remaining three observations (on 2021-04-14, 2021-05-10, and 2021-05-16) were taken with the B400/3400 grism and the R400/8500 grating.   Weather conditions were generally good except for the observation on 2021-05-10, which was taken under clear skies but very poor seeing (2--3$\arcsec$).   Because of the different readout times, the exposure durations on LRIS vary between the red and blue sides; in Table \ref{tab:spec_tab} we represent the individual exposures with average exposure time (over all exposures on both sides) for simplicity.

All spectra were reduced with LPipe \citep{Perley2019b}.
The two LRIS spectra in May do not show any discernible trace from the SN in the two-dimensional (2D) frames.  For the spectrum taken on 2021-05-16, we determine the position of the SN along the slit via its offset from the host nucleus (this slit was oriented across the nucleus at a PA of $-50^\circ$) and extract the flux at this location.  We also separately extract the flux of the entire host galaxy along the slit for spectroscopic analysis of the host.  (For the observation on 2021-05-10, the seeing is too poor to attempt separate site and host extractions, so we simply extract the host, but we do not use this spectrum in our subsequent analysis.)

\subsection{Multiwavelength}
\label{sec:multiwavelength}

\subsubsection{Swift/XRT}

We observed the field with \emph{Swift}'s onboard X-ray Telescope (XRT; \citealt{Burrows2005a}) in photon-counting mode, simultaneously with each set of UVOT observations (\S~\ref{sec:uvot}). There is no detection of the SN in any of these observations.  
Using the online tool\footnote{ \url{http://www.swift.ac.uk/user_objects/}}
provided by the UK \emph{Swift} team \citep{Evans2007a, Evans2009a}, 
we infer a median upper limit of $\sim0.006\,{\rm ct\,s}^{-1}$ per epoch at $3\sigma$ confidence. Stacking all data decreases the upper limit to $0.0008\,{\rm ct\,s}^{-1}$. Assuming a Galactic neutral hydrogen column density of $n(H)=2.4\times10^{20}\,{\rm cm}^{-2}$ \citep{HI4PI2016a} and a power-law spectrum with a photon index of 2, the count rates correspond to an unabsorbed flux limit of $2.2\times10^{-13}$ (for the median visit) and $3.1\times10^{-14}~{\rm erg\,cm}^{-2}\,{\rm s}^{-1}$ (for the stacked observation) in the bandpass 0.3--10\,keV. At the distance of SN\,2021csp, this corresponds to luminosity $L < 3.8\times10^{42}\,{\rm erg\,s}^{-1}$ and $L < 5.4\times10^{41}\,{\rm erg\,s}^{-1}$ at 0.3--10\,keV, respectively.

\subsubsection{Very Large Array}

We obtained three epochs of Very Large Array (VLA) observations of SN\,2021csp: one each on 2021-02-17, 2021-03-10, and 2021-07-10\footnote{Program IDs 20B-205 and 21A-308; PI Ho} using the X-band receiver.  The correlator was set up in 3-bit mode with frequency coverage from 8\,GHz to 12\,GHz.  In each observation the phase calibrator was J1430+1043 and the flux calibrator was 3C286.
Data were calibrated using the automated pipeline available in the Common Astronomy Software Applications (CASA; \citealt{McMullin2007}) and additional flagging was performed manually.
Data were imaged using the {\tt clean} algorithm \citep{Hogbom1974} with a cell size 1/10 of the synthesized beamwidth, and a field size of the smallest magic number
($10 \times 2^n$) larger than the number of cells needed to cover the primary beam.
The three observations all resulted in no detection of the counterpart, with root-mean-square (RMS) values of 5\,$\mu$Jy, 5\,$\mu$Jy, and 7\,$\mu$Jy for the February, March, and July observations (respectively).  The equivalent 3$\sigma$ limits on the source luminosity ($L_\nu$) are $2.6\times10^{27}$\,erg\,s$^{-1}$\,Hz$^{-1}$ at 7 and at 26 rest-frame days post-explosion, and $3.6\times10^{27}$\,erg\,s$^{-1}$\,Hz$^{-1}$ at 104 rest-frame days post-explosion.

\subsubsection{High-Energy Counterpart Search}

We searched the \emph{Fermi} Gamma-ray Burst Monitor (GBM) Burst Catalog \citep{NarayanaBhat2016}\footnote{\url{https://heasarc.gsfc.nasa.gov/W3Browse/fermi/fermigbrst.html}}, the \emph{Fermi}-GBM Subthreshold Trigger list\footnote{\url{https://gcn.gsfc.nasa.gov/fermi gbm\_subthresharchive.html}}, the \emph{Swift} Gamma-Ray Burst (GRB) Archive\footnote{\url{https://swift.gsfc.nasa.gov/archive/grb\_table/}}, the Interplanetary Network master list\footnote{\url{http://ipn3.ssl.berkeley.edu/masterli.txt}}, and the Gamma-Ray Coordinates Network archives\footnote{\url{https://gcn.gsfc.nasa.gov/gcn\_archive.html}} for a GRB between the last ZTF nondetection and the first ZTF detection. The closest event was one \emph{Fermi} burst (GRB\,210210B) $16^\circ$ away, but the association is unlikely given the size of the localization region. There was one IceCube event in the relevant time interval, but owing to the $10^\circ$ separation we consider the association unlikely.

\subsection{Host-Galaxy Photometry}

We retrieved science-ready coadded images from SDSS Data Release 9 
\citep{Ahn2012a}, UKIRT Infrared Deep Sky Survey DR11Plus \citep{Lawrence2007a}, and preprocessed \wise~\citep{Wright2010a} images from the unWISE archive \citep{Lang2014a}. The unWISE images are based on the public \wise\ data and include images from the ongoing NEOWISE-Reactivation mission R3 \citep{Mainzer2014a, Meisner2017a}. In addition to this, we use the UVOT observations that were obtained either before the explosion of SN\,2021csp or after the SN had faded from visibility. The brightness in the UVOT filters was measured with UVOT-specific tools in HEAsoft. Source counts were extracted from the images using a region of radius $10''$. The background was estimated using a circular region with a radius of $33''$ close to the SN position but not overlapping it. Count rates were obtained from the images using {\tt uvotsource}. They were converted to AB magnitudes using the UVOT calibration file from September 2020. 

We measured the brightness of the host using LAMBDAR 
\citep{Wright2016a}, {\tt uvotsource}, and the methods described by \citet{Schulze2021}. Table~\ref{tab:hostphot} provides the measurements in the different bands. 

\begin{deluxetable}{cccc}
\tablewidth{0pt}
\tablecaption{Photometry of the SN\,2021csp host galaxy\label{tab:hostphot}}
\tablehead{
\colhead{Survey/}   & \colhead{Filter} & \colhead{Magnitude} \\
\colhead{Telescope} & \colhead{}                & \colhead{}
}
\startdata
Swift/UVOT      &$ UVW2 $&$ 20.52 \pm 0.16 $ \\
Swift/UVOT      &$ UVM2 $&$ 20.47 \pm 0.07 $ \\
Swift/UVOT      &$ UVW1 $&$ 20.08 \pm 0.10 $ \\
SDSS            &$ u    $&$ 19.51 \pm 0.11 $ \\
SDSS            &$ g    $&$ 18.59 \pm 0.03 $ \\
SDSS            &$ r    $&$ 18.11 \pm 0.02 $ \\
SDSS            &$ i    $&$ 17.86 \pm 0.03 $ \\
SDSS            &$ z    $&$ 17.64 \pm 0.10 $ \\
PS1             &$ g    $&$ 18.47 \pm 0.04 $ \\
PS1             &$ r    $&$ 18.06 \pm 0.03 $ \\
PS1             &$ i    $&$ 17.83 \pm 0.05 $ \\
PS1             &$ z    $&$ 17.75 \pm 0.06 $ \\
PS1             &$ y    $&$ 17.59 \pm 0.17 $ \\
UKIDSS           &$ J    $&$ 17.66 \pm 0.03 $ \\
UKIDSS           &$ H    $&$ 17.57 \pm 0.07 $ \\
\textit{WISE}   &$ W1   $&$ 18.06 \pm 0.05 $ \\
\textit{WISE}   &$ W2   $&$ 18.56 \pm 0.06 $ \\
\enddata
\tablecomments{All magnitudes are reported in the AB system and not corrected for extinction.}
\end{deluxetable}

\section{Analysis} \label{sec:analysis}

\subsection{Light Curve}

\subsubsection{Explosion Time}
\label{sec:exptime}

SN\,2021csp was identified prior to peak brightness and upper limits shortly prior to discovery are available, permitting a reasonably tight constraint on the time of first light (defined here as the moment when optical photons in excess of the progenitor luminosity are first able to escape and travel freely toward the observer).  We will refer to this as the ``explosion time'' for simplicity, although we emphasize that the data cannot actually separately distinguish the time of core collapse or shock breakout.

The most recent ZTF/P48 upper limit prior to the discovery is from an observation at MJD 59254.52578 ($g > 21.50$\ mag, 2.5$\sigma$), which is 1.94\ days before the first detection in the $i$ band and 1.97\ days prior to the first detection in the $g$ band.   Assuming an early-time flux evolution following $F \propto t^2$, the earliest explosion time consistent with the $g$-band limit is ${\rm MJD}_{\rm exp} > 59254.0$.   
This limit is likely to be conservative, since the flux was already turning over from a $t^2$-like early behavior at the time of the initial detections.  

No upper limit can be formally placed on the time of explosion other than the time of the first detection itself since the rising phase is too short and poorly sampled to be modeled effectively.  The (very conservative) upper limit is thus ${\rm MJD}_{\rm exp} < 59256.47$.  Given that the source was already quite bright at this time, our general expectation (supported by the blackbody modeling; \S \ref{sec:blackbody}) is that the explosion time is probably closer to the beginning of the constrained window.

Throughout the remainder of the paper we will express observation times in the rest frame relative to MJD 59254.5, the approximate time of the last upper limit and a reasonable guess of the time of explosion.  
Expressed in this system, our constraint on the actual time of explosion is $-0.46$\,d $< t_{\exp} < 1.82$\,d.

\subsubsection{Characteristic Timescale}

To better quantify the rapid evolution of SN\,2021csp and compare it to other optical transients, we perform a basic measurement of the characteristic evolutionary timescales. 

The rise time ($t_{\rm rise}$, defined as the rest-frame time from explosion to peak brightness) depends on the band, with redder filters showing later peaks (and therefore longer rise times).  In the (observed) $g$ band where the early light curve is best sampled, the rise time is 1.8--4.0 rest-frame days, with the large uncertainty originating primarily from the uncertainty in the explosion time itself (although following the arguments in \S\,\ref{sec:exptime}, times toward the upper end of this range are likely more plausible).  The rise time is $\sim 1$\ day longer in $r$ and $\sim1.5$\ days longer in $i$ and $z$. 

For comparison to the light curves of other SNe, a standard metric is the half-maximum time ($t_{1/2}$), the amount of time (rest frame) which the transient spends at a flux level more than half of its maximum in some wavelength band.  This can be decomposed into separate half-rise ($t_{\rm 1/2,rise}$) and half-fade ($t_{\rm 1/2,fade}$) times, the intervals over which the transient rises from half-maximum to maximum brightness and fades from maximum to half-maximum brightness (respectively). The smoothed interpolation of our $g$-band light curve gives a half-rise time of $t_{\rm 1/2,rise}=2.5\pm0.5$ days and a half-fade time of $t_{\rm 1/2,fade}=8.3\pm1$ days, for a total time above half-max of $t_{1/2} = 10.8\pm1.2$ days.  (The $r$-band timescale is somewhat slower, with $t_{1/2} \approx 15$\ days).

\begin{figure*}[ht]
\centering
\includegraphics[width=\linewidth]{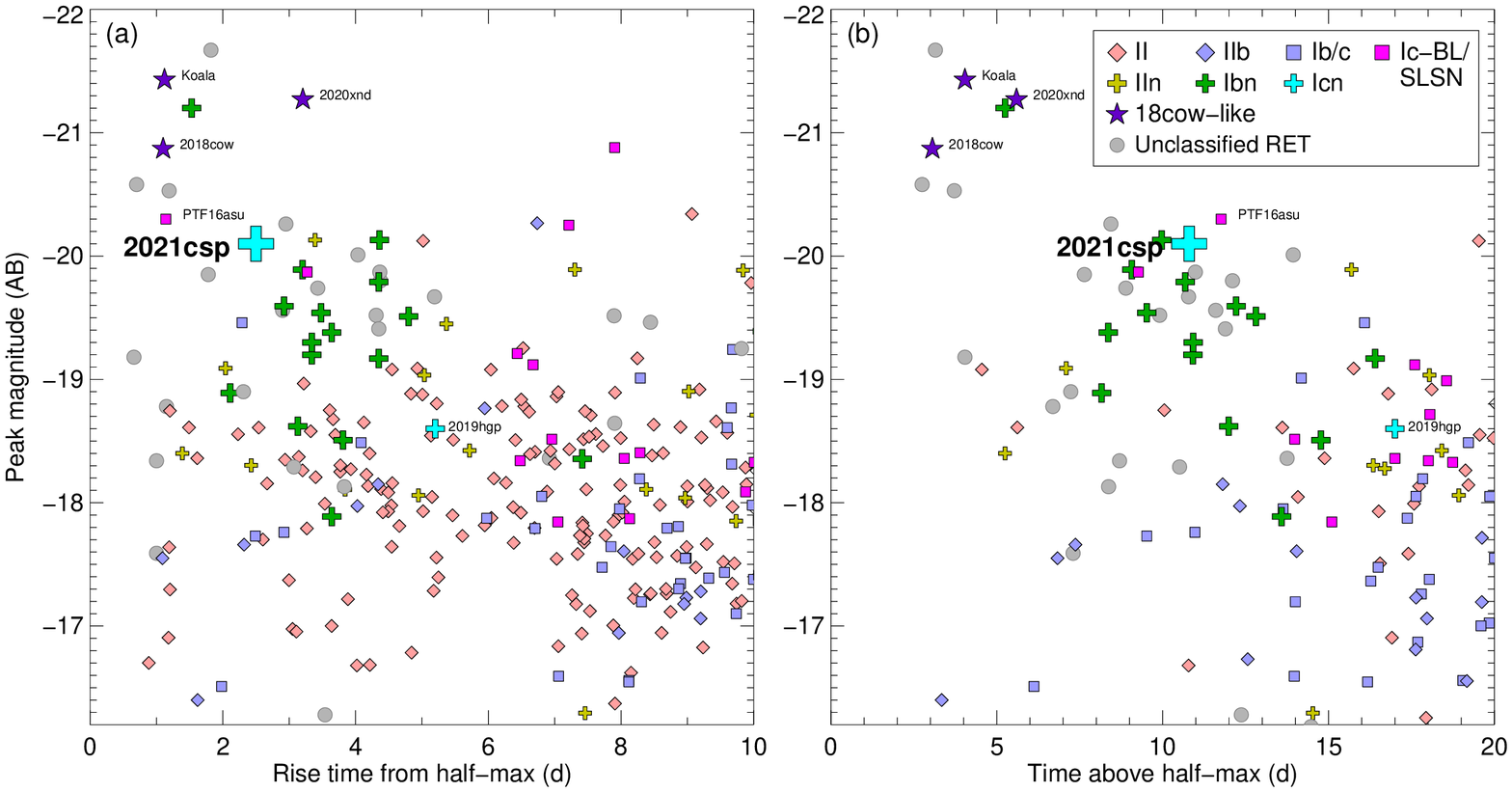}
\caption{Characteristic timescales for SN\,2021csp compared to the known population of core-collapse transients from the ZTF Bright Transient Survey \citep{Perley2020} and to fast transients ($t_{1/2} < 12$\ days) from the literature \citep{Drout2014,Pursiainen2018,Ho2019gep,Ho2021}.
Panel~(a) at left shows the rise time from half-maximum to maximum brightness ($t_{\rm 1/2,rise}$) on the abscissa; panel~(b) at right shows the total time above half-maximum ($t_{1/2}$) on the abscissa.  The properties of SN\,2021csp are similar to those of SNe~Ibn, although it shares some features with the AT\,2018cow-like population.  An earlier SN~Icn, SN\,2019hgp, is less extreme but also overlaps the SN~Ibn distribution.}
\label{fig:timescale}
\end{figure*}

A comparison between the characteristic timescales and luminosities of SN\,2021csp and similarly-measured estimates for a variety of other ``fast'' transients is shown in Figure~\ref{fig:timescale}.  SN\,2021csp is much more extreme than SN\,2019hgp and fits in well with the population of spectroscopically-unclassified fast and luminous optical transients from the works of \cite{Drout2014} and \cite{Pursiainen2018} (gray circles).

More recently, \cite{Ho2021} compiled a large catalog of rapidly-evolving events with spectroscopic classifications from the ZTF high-cadence surveys (1\ day cadence or faster), and \cite{Perley2020} produced a spectroscopically-complete catalog of events from the ZTF public Bright Transient Survey (3\ day cadence).  The samples from these two surveys are added to Figure~\ref{fig:timescale} for comparison.  Consistent with its spectroscopic properties, SN\,2021csp is sited in the same region of parameter space occupied by interaction-dominated transients (primarily SNe~Ibn and fast SNe~IIn; see \citealt{Ho2021}).  However, it is among the most luminous examples of this group and also one of the fastest-rising, bringing it closer to the ``Cow-like'' radio-loud population in the top left of Figures \ref{fig:timescale}a-b.

\subsubsection{Blackbody modeling}
\label{sec:blackbody}

\begin{figure*}
    \centering
    \includegraphics[width=0.97\textwidth]{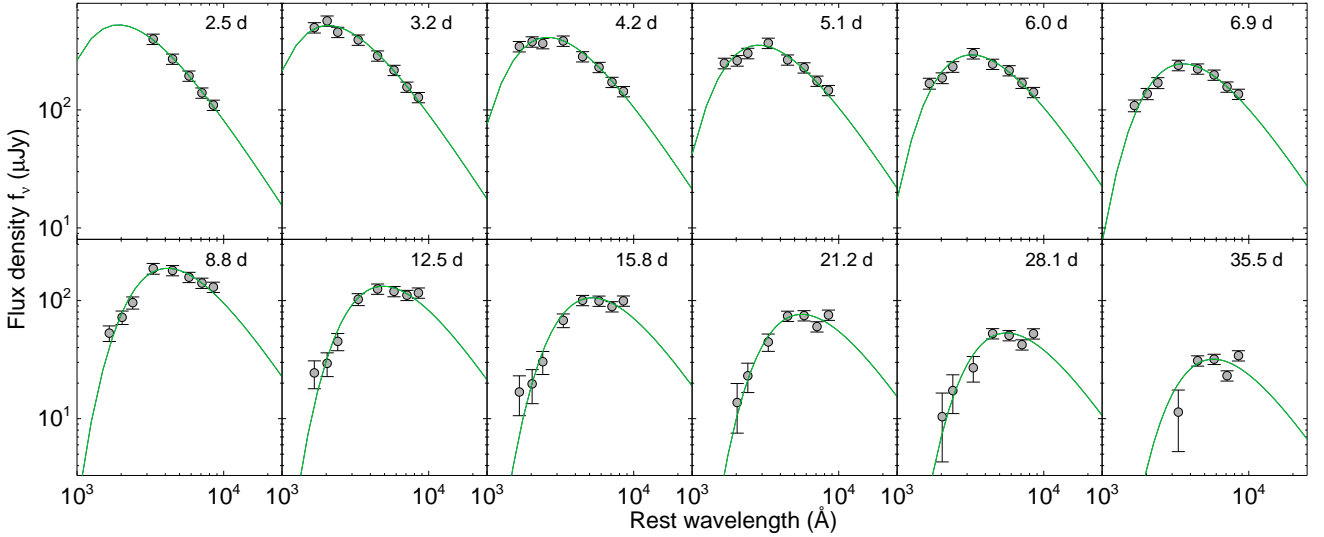}
    \caption{Blackbody fits to interpolated SEDs at various post-explosion rest-frame times.  Error bars show 2$\sigma$ uncertainties.
    }
    \label{fig:bbfits}
\end{figure*}

To obtain common-epoch spectral energy distributions (SEDs), we define a set of standardized epochs (chosen to be close in time to actual multiband measurements) and use a combination of local regression smoothing and spline fitting to obtain interpolated light-curve measurements for all available filters at each point.  After correcting for Galactic extinction, we then fit a Planck function to each set of fluxes to determine the effective temperature, photospheric radius, and luminosity.  The host-galaxy extinction $E_{B-V,{\rm host}}$ is initially assumed to be zero (based on the
face-on geometry of the host, the outlying location of the event, and the lack of narrow absorption lines from the interstellar medium in the spectra), but we later repeat the procedure under different assumptions about the host reddening.  The SED fits are shown in Figure \ref{fig:bbfits}.

\begin{figure}
    \centering
    \includegraphics[width=0.47\textwidth]{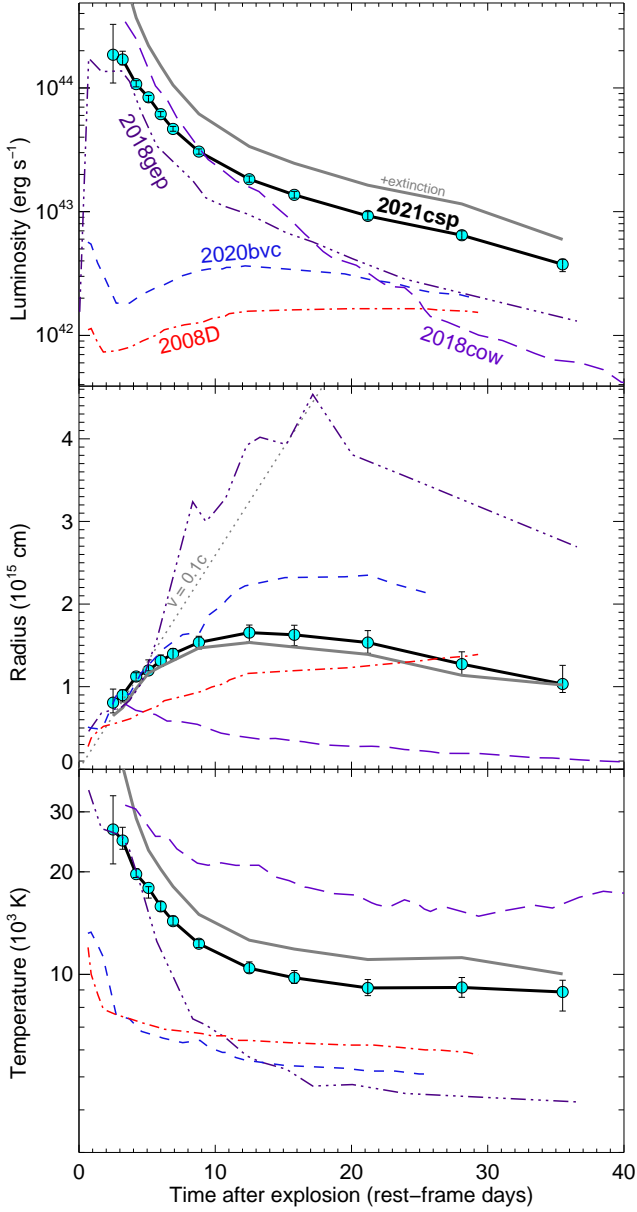}
    \caption{Evolution of photospheric parameters estimated from blackbody fits to the UV-optical SED of SN\,2021csp during the first month.  The solid black curves show results assuming no host-galaxy extinction; the solid gray curves assume $E_{B-V,{\rm host}}=0.15$\ mag.  Various comparison objects with fast early evolution from the literature are shown for comparison: SN\,2008D, a ``normal'' SN~Ib \citep{Modjaz2009}, SN\,2020bvc, a ``normal'' SN~Ic-BL \citep{Ho2020bvc}, SN\,2018gep (a strongly interacting SN~Ic-BL; \citealt{Ho2019gep,Pritchard2021}), and AT\,2018cow (an extreme FBOT which did not develop any late-time SN; \citealt{Perley2019}).  The radii of all of these explosions are similar at $\sim 5$\ days post-explosion, indicating similar ejecta velocities ($v \approx 0.1$c).  Different evolution sets in at later phases.  (Note: the late-time rapid downturn is not shown here owing to the lack of UV photometry to constrain the temperature after 35\ days.)
    }
    \label{fig:physevol}
\end{figure}

The physical parameters (blackbody luminosity, radius, and temperature) derived from these fits are shown in Figure~\ref{fig:physevol} (and provided in Table \ref{tab:physevol}), where they are compared with a variety of other fast and/or luminous transients measured using similar approaches.    The fast rise to peak brightness implies an initial velocity that is very high ($\sim$30000\,km\,s$^{-1}$), akin to what is seen in SNe~Ic-BL. 

\begin{deluxetable}{ccccc}
\tablewidth{0pt}
\tablecaption{Results of blackbody modeling\label{tab:physevol}}
\tablehead{
\colhead{MJD}   & \colhead{$t$} & \colhead{log$_{10}$($L$)} & \colhead{log$_{10}$($R$)} & \colhead{log$_{10}$($T$)} \\
\colhead{} & \colhead{(d)} & \colhead{(erg\,s$^{-1})$} & \colhead{(cm)} & \colhead{(K)}
}
\startdata
59257.20 &   2.49 & $44.33^{+0.17}_{-0.28}$ & $14.88^{+0.10}_{-0.06}$ & $ 4.45^{+0.07}_{-0.12}$ \\ 
59258.00 &   3.23 & $44.17^{+0.15}_{-0.02}$ & $14.98^{+0.01}_{-0.07}$ & $ 4.37^{+0.07}_{-0.00}$ \\ 
59259.00 &   4.15 & $44.00^{+0.04}_{-0.03}$ & $15.06^{+0.02}_{-0.03}$ & $ 4.28^{+0.02}_{-0.01}$ \\ 
59260.00 &   5.07 & $43.86^{+0.04}_{-0.02}$ & $15.12^{+0.02}_{-0.03}$ & $ 4.21^{+0.02}_{-0.01}$ \\ 
59261.00 &   6.00 & $43.74^{+0.04}_{-0.02}$ & $15.15^{+0.03}_{-0.03}$ & $ 4.17^{+0.02}_{-0.02}$ \\ 
59262.00 &   6.92 & $43.62^{+0.03}_{-0.02}$ & $15.18^{+0.02}_{-0.03}$ & $ 4.13^{+0.02}_{-0.01}$ \\ 
59264.00 &   8.76 & $43.46^{+0.02}_{-0.02}$ & $15.21^{+0.03}_{-0.03}$ & $ 4.07^{+0.02}_{-0.01}$ \\ 
59268.00 &  12.45 & $43.23^{+0.03}_{-0.02}$ & $15.25^{+0.03}_{-0.03}$ & $ 4.00^{+0.02}_{-0.02}$ \\ 
59271.00 &  15.22 & $43.13^{+0.03}_{-0.03}$ & $15.25^{+0.03}_{-0.03}$ & $ 3.97^{+0.02}_{-0.02}$ \\ 
59285.00 &  28.14 & $42.80^{+0.04}_{-0.03}$ & $15.10^{+0.04}_{-0.04}$ & $ 3.96^{+0.02}_{-0.02}$ \\ 
59293.00 &  35.52 & $42.55^{+0.06}_{-0.04}$ & $15.04^{+0.05}_{-0.06}$ & $ 3.93^{+0.05}_{-0.03}$ \\ 
\enddata
\tablecomments{Uncertainties are statistical only and do not include systematic effects associated with the unknown host extinction ($A_V = 0$ is assumed) or deviations of the true spectral shape from a single-temperature blackbody.}
\end{deluxetable}

The subsequent evolution is generally normal, in the sense that the luminosity and temperature decline while the photospheric radius increases, reaches a maximum, and then recedes into the cooling ejecta.  The final two points should be treated with caution, since at this point the spectrum has heavily diverged from a simple blackbody (Fig.~\ref{fig:optspec}) and the UV emission is weak or absent.

We examined whether the possibility of host extinction would alter any of the above conclusions.  For a Milky Way-like reddening law (\citealt{Fitzpatrick1999}), the maximum potential extinction permitted by our SED models is $E_{B-V,{\rm host}} = 0.15$\ mag (higher extinction values lead to poor fits at early times because the corrected fluxes become too blue for a blackbody model.)  The inferred luminosity and temperature both increase significantly at early times in this scenario, but the radius measurements are affected by only 10--20\% (see dotted lines in Fig.~\ref{fig:physevol}).  For the remainder of the discussion we will continue to assume $E_{B-V,{\rm host}} = 0$.

\subsection{Spectral analysis}

The spectroscopic sequence in Figure~\ref{fig:optspec} shows two distinct regimes. At 2--10\ days, the spectra are characterized by a hot, blue continuum superimposed with strong, narrow P~Cygni features (``narrow phase'').   After 16\ days, the narrow lines have disappeared completely and a series of broad features with velocities characteristic of SN ejecta emerge instead (``broad phase'').  The spectrum in between these two periods (i.e., 10--16 days) exhibits a brief transitional state in which most of the narrow optical features have vanished \emph{but} \ion{C}{2} remains and the UV P~Cygni features also remain very strong, and whereas broad features are becoming evident in the spectrum they are still weak and indistinct.  We summarize the key features of the two spectral regimes below.

\subsubsection{Narrow-phase spectra}

\begin{figure*}
    \centering
    \includegraphics[width=0.97\textwidth]{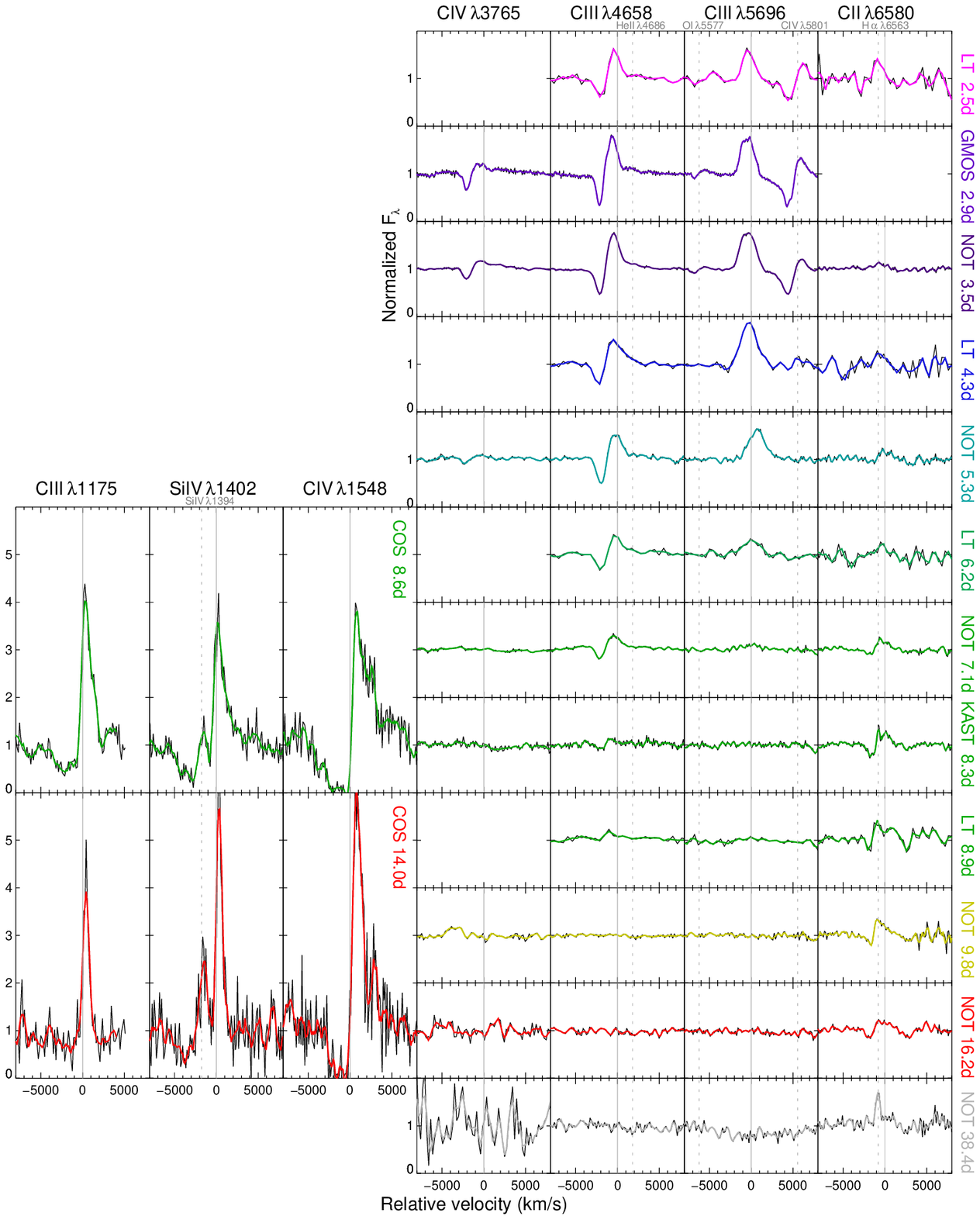}
    \caption{Evolution of selected narrow-line features in the UV (left three columns) and optical (right four columns).  Original spectra are plotted in black; a smoothing kernel has been applied to the colored curves.  Line centers (assuming $z=0.084$) are indicated as solid lines with other nearby (contaminating) transitions indicated as dotted lines.  The \ion{C}{4}\,$\lambda3765$ and $\lambda5801$ features disappear from the optical spectra at $\sim 5$\ days; the \ion{C}{3}\,$\lambda$4658 and $\lambda$5696 features disappear at $\sim 9$\ days.  Weak \ion{C}{2}\,$\lambda$6580 persists until later times, although it becomes contaminated by host-galaxy H$\alpha$ emission and is difficult to recognize after 16\ days.  The UV features remain strong until at least 14\ days.  (All times refer to rest-frame elapsed time following the assumed explosion epoch.)
    }
    \label{fig:speczoom}
\end{figure*}

All identified strong lines spanning the UV to 8000\,\AA\ are shown in 
Figure~\ref{fig:uvoptspec} (and listed in Table \ref{tab:lines}), with close-up views of various strong features presented in Figure~\ref{fig:speczoom}. 
Almost all of the identifiable lines are associated with oxygen, carbon, silicon, or magnesium.  \ion{He}{2} $\lambda$4686 may be present in a blend with the \ion{C}{3} $\lambda$4656 feature, although because of the high velocities this cannot be 
conclusively established.
However, \ion{He}{1}\,$\lambda$5876 
is clearly seen.  Some of the later spectra show a P~Cygni feature close to the position of H$\alpha$\,$\lambda$6563, although more likely this feature originates from a combination of \ion{C}{2}\,$\lambda$6580 (which persists longer than the other lines) and host-galaxy narrow emission.  Most line profiles have a P~Cygni shape, with blueshifted absorption and emission that may be either net blueshifted or net redshifted depending on the line and phase.  The far-UV Si lines are seen only in emission, as is \ion{C}{3}\,$\lambda$5696.

\begin{figure*}
    \centering
    \includegraphics[width=0.97\textwidth]{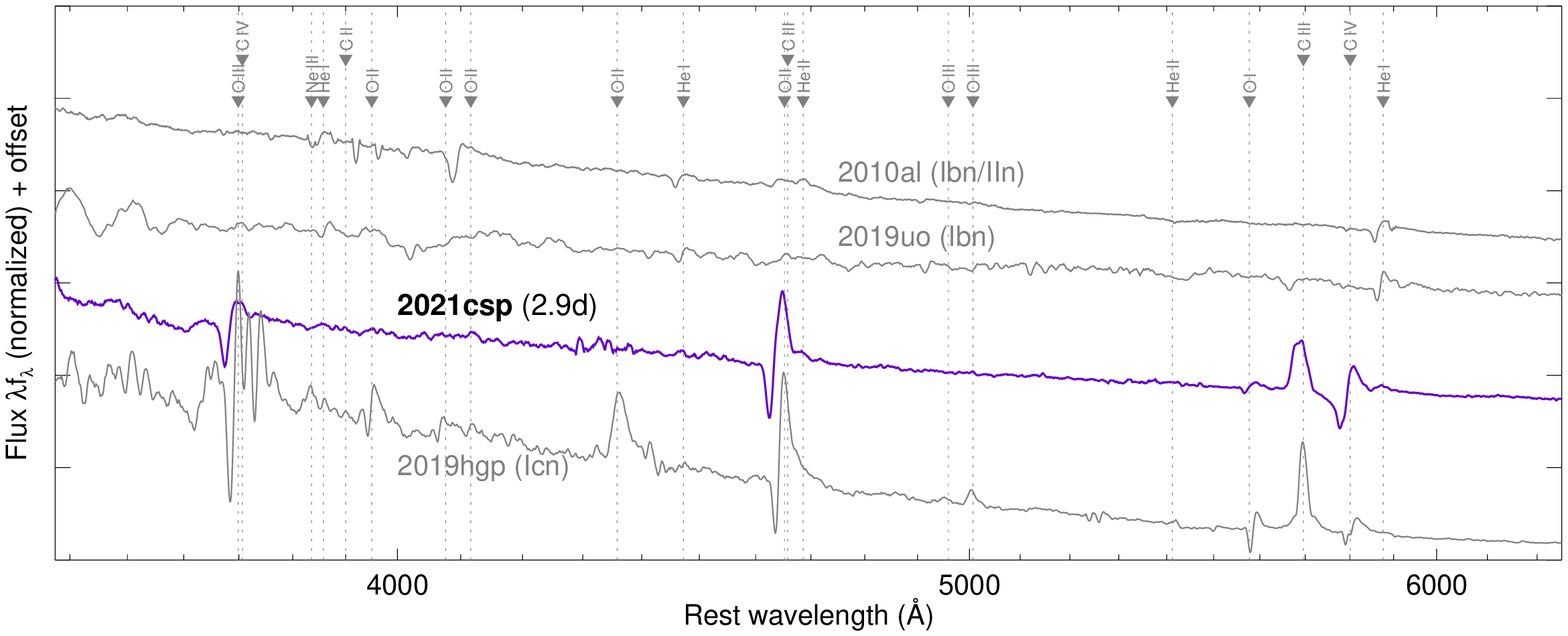}
    \caption{Peak-light spectra of SN\,2021csp and SN\,2019hgp (Type Icn) and of SN\,2019uo and SN\,2010al (Type Ibn).  Both Type Ibn and Icn SNe show narrow P~Cygni features in their early-time spectra, mostly from He for Ibn SNe and C for Icn SNe.  The lines are much stronger in the SN~Icn spectra.
    }
    \label{fig:specvs19hgp}
\end{figure*}

\begin{figure*}
    \centering
    \includegraphics[width=0.97\textwidth]{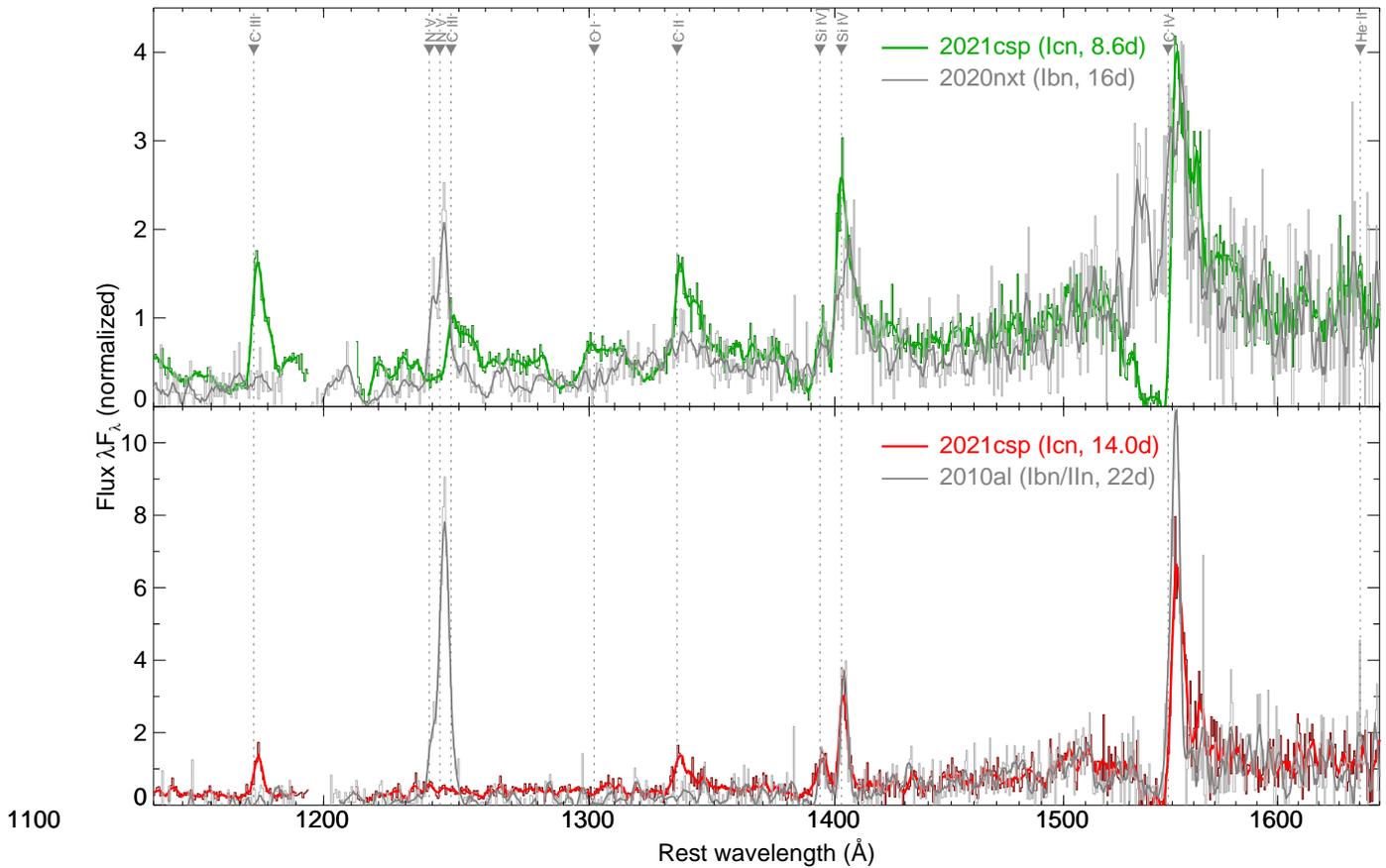}
    \caption{UV spectra of SN\,2021csp compared with those of two SNe~Ibn.  The early-time spectrum of SN\,2021csp is compared to the Type Ibn SN\,2020nxt (Fox et al. 2021, in prep.); the later spectrum is compared to the (later-phase) observation of the Type Ibn SN\,2010al \citep{Kirshner+2010}.
    }
    \label{fig:specuvcompare}
\end{figure*}

\begin{deluxetable}{cl}
\tablewidth{0pt}
\tablecaption{Spectral Lines\label{tab:lines}}
\tablehead{
\colhead{Line}   & \colhead{Rest wavelength} \\
\colhead{} & \colhead{(\AA)}
}
\startdata
C~II   & 1036 \\
Si~IV  & 1062 \\
Si~IV  & 1073 \\
C~III  & 1175.7  \\
C~III  & 1247.38 \\
O~I\tablenotemark{a}    & 1302 \\ 
C~II\tablenotemark{a}   & 1335 \\ 
Si~IV  & 1393.76 \\
Si~IV  & 1402.77 \\
C~IV   & 1548 \\
Al~III & 1854.73 \\
Al~III & 1862.79 \\
C~III  & 1908 \\
C~III  & 2010.1 \\
C~III  & 2296.87 \\
C~II   & 2324 \\
C~IV   & 2404.77 \\
He~II  & 2511.4 \\
Mg~II  & 2800 \\
C~IV   & 3765 \\
C~III  & 4658 \\
He~II  & 4686 \\
O~I    & 5577 \\
C~IV   & 5801 \\
C~III  & 5695.92 \\
He~I   & 5876 \\
C~II   & 6580 \\
C~II   & 7234 \\
\enddata
\tablecomments{Narrow (CSM) lines identified in the UV and optical spectra of SN\,2021csp and shown in Figures.}
\tablenotetext{a}{Uncertain line association}
\end{deluxetable}

Despite being qualitatively characterized as narrow lines, the velocities inferred from these features are quite high.  The deepest point of absorption in the strong lines from the early, high-S/N optical spectra is at $-2200$\,km\,s$^{-1}$, with a maximum blueshift (blue edge) of $-3000$\,km\,s$^{-1}$.  The inferred velocities in the UV (where the transitions are much stronger) are even higher; the \ion{C}{4}\,$\lambda$1548 line shows almost-total absorption out to $-2000$\,km\,s$^{-1}$ but weaker absorption out to a maximum blueshift of approximately $-4500$\,km\,s$^{-1}$.

A comparison between the peak-light spectra of SN\,2021csp, the prototypical Type Icn SN\,2019hgp, and two SNe~Ibn (SN\,2019uo and SN\,2010al) is displayed in Figure~\ref{fig:specvs19hgp}.  The spectrum of SN\,2019uo is the classification spectrum from the Transient Name Server \citep{TNSCR188}; the spectrum of SN\,2010al is taken from \cite{Pastorello2015c}.  The spectrum of SN\,2021csp strongly resembles that of SN\,2019hgp, although it lacks some of the transitions seen in that SN (e.g., \ion{O}{3}).  The line widths in SN\,2019hgp are somewhat broader.  The features in the SNe~Ibn originate from different transitions (mainly \ion{He}{1}) and are much weaker than those of either SN~Icn, although the line profiles are qualitatively similar.

A comparison with two SNe~Ibn in the UV (SN\,2020nxt and SN\,2010al; Fox et al.\ 2021, in prep.; \citealt{Kirshner+2010}) is provided in Figure~\ref{fig:specuvcompare}.  
Some common transitions are apparent at these wavelengths, most notably the resonance lines of \ion{Si}{4}\,$\lambda$1402 and \ion{C}{4}\,$\lambda$1548, which have similar strengths and profiles.  The remaining features are quite different: SN\,2021csp shows a number of carbon features absent in SNe~Ibn, while the very strong doublet \ion{N}{5}\,$\lambda\lambda$1238,1242 is seen in both SNe~Ibn but is absent entirely in the Type Icn SN\,2021csp.
Also, while the characteristic velocities are similar, the high-velocity component (4000\,km\,s$^{-1}$) in absorption and emission seen in SN\,2021csp is not clearly visible in either of the SNe Ibn --- although the issue is somewhat confused by contamination with other features and the different phases of the observations.

\subsubsection{Broad-phase spectra}

The broad lines are somewhat indistinct at 10--15\ days, but by 16\ days the characteristic late-time spectrum has clearly emerged.  The flux is strongest in the blue, with maxima at 4600\,\AA, $\sim 5300$\,\AA, and $\sim 6400$\,\AA.  
The relative strength of these features increases
gradually with time, but neither their shapes nor central wavelengths change much.  A notable exception is the \ion{Ca}{2} near-infrared triplet at $\sim 8540$\,\AA:  not apparent at all prior to $\sim30$\ days, it rapidly rises to become the dominant emission feature in our final spectrum at 53\ days.   The maximum velocity (at zero intensity) of this feature on the blueshifted side is $\sim -10000$\,km\,s$^{-1}$, characteristic of nebular-phase stripped-envelope SNe.

\begin{figure*}
    \centering
    \includegraphics[width=0.95\textwidth]{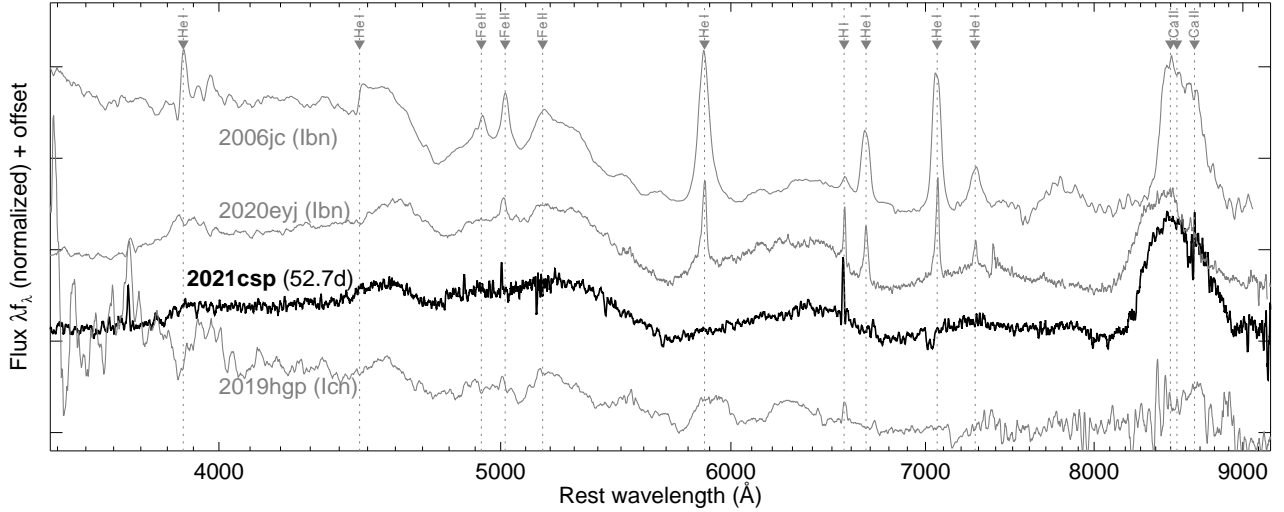}
    \caption{The late-time spectrum of SN\,2021csp (from Keck/LRIS) compared to Type Icn SN\,2019hgp \citep{GalYam2021} and two SNe~Ibn (SN\,2006jc from \citealt{Pastorello2007} and SN\,2020eyj from Kool et al. 2021, in prep.).
    }
    \label{fig:specvsIbn}
\end{figure*}

The identity of the remaining features is less clear.  While the late-phase spectra show some similarities to those of SNe Ic, the strong blue and near-UV continuum differs dramatically from the line-blanketed post-peak spectra of any normal member of this class (including SNe Ic-BL).  Instead, the continuum strongly resembles those of SNe~Ibn at similar phases, although the narrow \ion{He}{1} lines characteristic of SNe~Ibn at these phases are absent. 
A comparison between SN\,2021csp, SN\,2019hgp, and two late-phase SNe~Ibn (SN\,2006jc from \citealt{Pastorello2007} and SN\,2020eyj from Kool et al. 2021, in prep.) is shown in Figure~\ref{fig:specvsIbn}.
The blue pseudocontinuum seen in SNe~Ibn has been attributed to a forest of blended \ion{Fe}{2} lines provided by fluorescence in the inner wind or post-shock gas \citep{Foley2007,Chugai2009,Smith2009,Pastorello2015}, and its presence here suggests that strong CSM interaction is continuing even after the narrow lines have faded.

\subsection{Polarimetry}

An upper limit on the interstellar 
polarization (ISP) induced by dichroic extinction of Milky Way-like dust grains 
is given by $p_{\rm ISP} < 9\times E_{B-V}$ \citep{Serkowski1975}. 
Therefore, we set an upper limit on the ISP from the Galactic component as 0.24\%. 
We assume a host $A_V = 0$\ mag (\S \ref{sec:blackbody}).
We evaluated a continuum polarization level of $\sim 0.3$\% 
by computing the error-weighted Stokes parameters in the optical range after 
excluding the prominent spectral features and telluric ranges. Therefore, 
without a careful determination of the ISP from the SN host, we suggest that 
the continuum polarization of the SN is less than $\sim 0.5$\%.

There is no strong polarization signal associated with any of the narrow-line features, although the wavelength bins in the vicinity of flash-ionized narrow P~Cygni features of ionized \ion{C}{3} and \ion{C}{4} (labeled in Fig.~\ref{fig:specpol}) do show a polarization excess of $\sim 0.4$\% above the continuum level at $\sim 5\sigma$ significance, which may be an indicator of some (limited) asymmetry in the explosion and/or CSM.

Assuming a limiting polarization of 0.5\%, 
the axis ratio of the photosphere can be limited to $\lesssim 1.3$ assuming an ellipsoidal surface with a Thomson optical depth of 5 and a radial CSM density profile of $n(r) \propto r^{-n}$, with an index $n$ in the range 3--5 \citep{Hoflich1991}.

\subsection{Radio Analysis}

\begin{figure}
    \centering
    \includegraphics[width=0.46\textwidth]{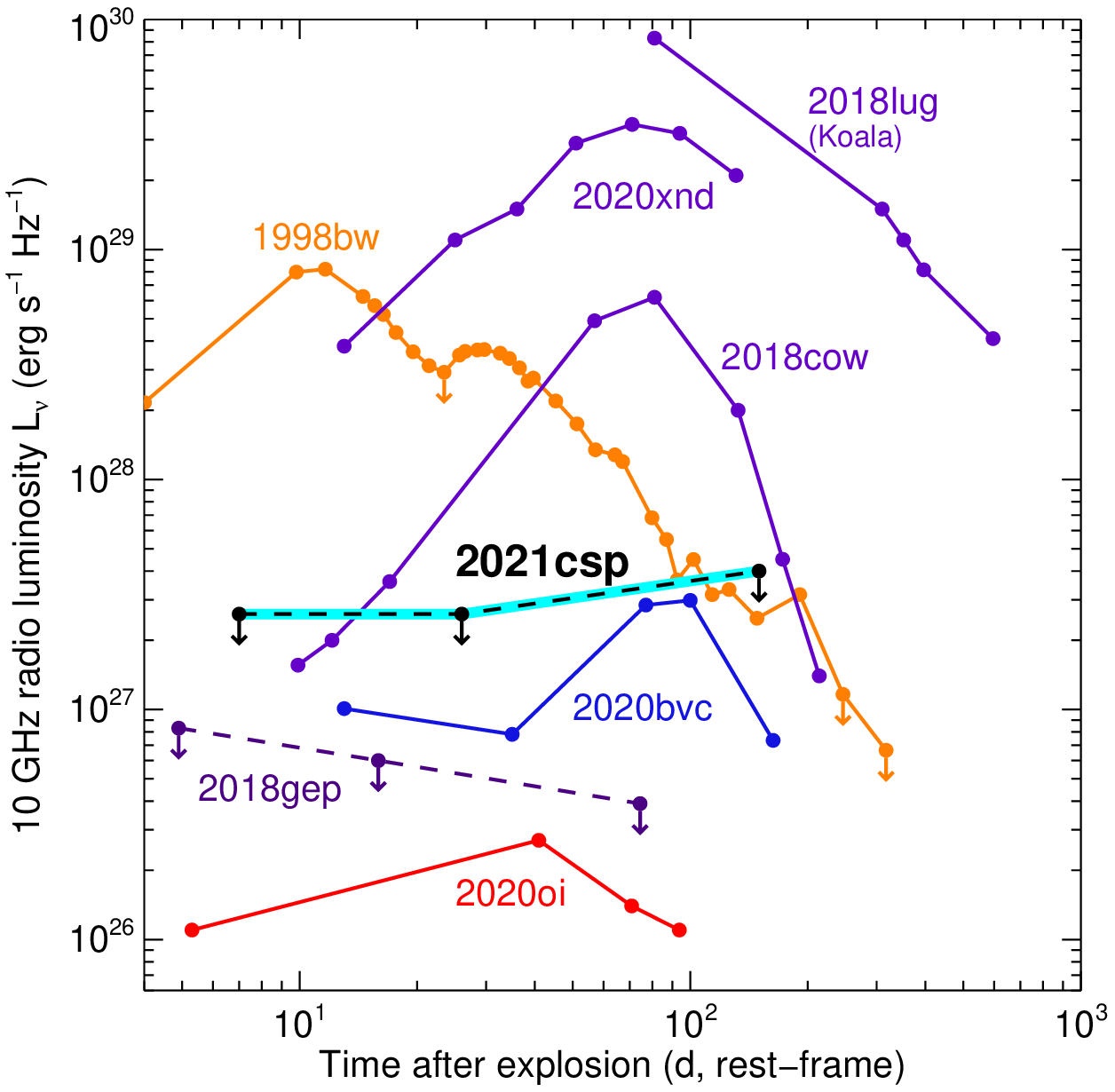}
    \caption{$3\sigma$ upper limits on the radio luminosity of SN\,2021csp, compared to rapidly-evolving transients reported by \cite{Ho2021} and other events from the literature: the radio-loud (``Cow-like'') FBOTs AT\,2018cow \citep{Margutti2019}, CSS\,161010 \citep{Coppejans2020}, AT\,2020xnd \citep{Ho2021xnd}, and AT\,2021lug \citep{Ho2020koala}; the GRB-associated broad-lined SN~Ic-BL 1998bw \citep{Kulkarni1998,Wieringa1999}; the ``normal'' Type Ic-BL SN\,2020bvc \citep[][and work in prep.]{Ho2020bvc}, the strongly-interacting Type Ic-BL SN\,2018gep \citep{Ho2019gep}; and the ``normal'' Type Ic SN\,2020oi \citep{Horesh2020}.
    A radio counterpart as luminous as that seen in radio-loud FBOTs  can be ruled out, but not a fainter source such as what was observed in SN\,2020oi or SN\,2020bvc.
    }
    \label{fig:radiolc}
\end{figure}

SN\,2021csp was not detected in any of our radio observations.  The radio limits do not rule out a light curve similar to that seen in ordinary SNe, but the luminosity limits derived from the second and third measurements are significantly below the light curves of AT\,2018cow or AT\,2020xnd at comparable epochs \citep{Ho2021}.  A comparison between the upper limits and some previous SN radio light curves is shown in Figure~\ref{fig:radiolc}.

\subsection{Host Galaxy}

\begin{figure}[ht]
\centering
\includegraphics[width=0.45\textwidth]{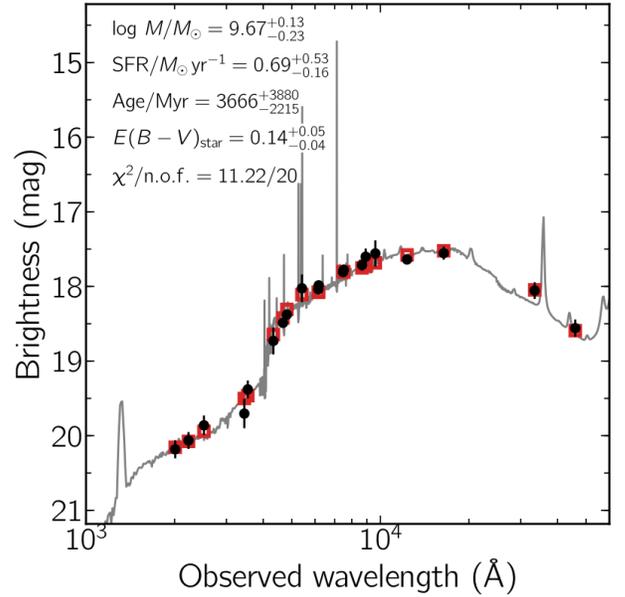} 
\caption{SED of the host galaxy of SN\,2021csp (black data points). The solid line displays the best-fitting model of the SED. The red squares represent the model-predicted magnitudes. The fitting parameters are shown in the upper-left corner. The abbreviation ``n.o.f.'' stands for number of filters.
}
\label{fig:gal_sed}
\end{figure}

We modeled the host-galaxy SED with the software package {\tt prospector} \citep{Leja2017a} using the procedures of \cite{Schulze2021}.
We assumed a Chabrier initial mass function \citep{Chabrier2003a}, approximated the star-formation history (SFH) by a linearly increasing SFH at early times followed by an exponential decline at late times (functional form $t \times \exp\left(-t/\tau\right)$), and used a \citet{Calzetti2000a} attenuation law.

Figure~\ref{fig:gal_sed} shows the observed SED and its best fit. The SED is adequately described by a galaxy template with a mass of $\rm{log}(M/M_\odot) = 9.67^{+0.13}_{-0.23}$ and a star-formation rate of $0.69^{+0.53}_{-0.16}\,M_\odot\,{\rm yr}^{-1}$.

Emission-line fluxes were extracted from the late-time Keck spectrum (using the observation from 2021-05-16, which covered the host nucleus and was taken after the transient had faded; we use a custom extraction covering the entire host).
We measure the following line fluxes for H$\alpha$, H$\beta$, [\ion{O}{3}]\,$\lambda$5007, [\ion{O}{3}]\,$\lambda$4959, and [\ion{N}{2}]\,$\lambda$6584 of $42.2\pm0.4$, $10.8\pm0.7$, $14.1\pm0.8$, $4.3\pm0.6$, and $11.0\pm0.4$, respectively (units of $\times10^{-16}\,{\rm erg\,cm}^{-2}\,{\rm s}^{-1}$; no extinction correction has been applied).
We estimate the metallicity at the galaxy center using the O3N2 indicator with the calibration reported by \cite{Marino2013a}. The oxygen abundance of $12+\log({\rm O/H})=8.35\pm0.01$ translates to a moderately low 
metallicity
of $(0.46\pm0.01)$\,$Z_\odot$
(assuming a solar oxygen abundance of 8.69; \citealt{Asplund2009a}). 

\subsection{Summary of Observational Properties}

The key observational features of SN\,2021csp are summarized below.

\begin{itemize}
  \item SN\,2021csp exhibits three distinct phases.  At early times ($<10$\ days), the temperature is very high but rapidly cooling, and the spectra are dominated by strong, narrow P~Cygni features of C and O.  At 20--60\ days, the spectra are dominated by broad features and there is comparatively little color evolution; the light curve declines gradually.  After 60\ days, the light curve fades very rapidly and the transient disappears (absolute magnitude $M_r > -13$) by 80\ days.
  \item The spectra are dominated by C and O, with Si also evident in the UV and an Fe pseudocontinuum visible in the broad-component phase in the blue.  Compared with SNe~Ibn, He is weak while N and H are absent. The strength of the narrow lines is greater than in any known SN~Ibn close to peak brightness, but narrow lines are lacking entirely at late times.
  \item Several characteristic velocities are evident.  The ``narrow'' features show maximum absorption at 2000\,km\,s$^{-1}$ with a maximum velocity of 4500\,km\,s$^{-1}$, indicative of the velocity of the CSM.  The early photospheric modeling indicates the existence of a high-velocity ejecta component with 30000\,km\,s$^{-1}$. Late-phase optical spectra suggest a characteristic ejecta velocity of 10000\,km\,s$^{-1}$. 
  \item The very fast rise (3\ days) and high peak luminosity ($M_g \approx -20$\ mag) are consistent with common definitions for an ``FBOT'', but these values are not unusual for SNe~Ibn, many of which have also been shown to be ``FBOTs'' \citep{Ho2021}.
  \item There is no detection of a radio or X-ray counterpart.  The limits rule out an AT\,2018cow-like event or GRB, but not most classes of normal SNe.
  \item The transient occurred in the outer regions of a moderately low-mass, star-forming spiral galaxy.
\end{itemize}

Key observational parameters are summarized for convenience in Table \ref{tab:properties}.

\begin{deluxetable}{lll}
    \tablecaption{Key properties of SN\,2021csp}
    \label{tab:properties}
\tablehead{
\colhead{Property}   & \colhead{Value} & \colhead{Description}
}
\startdata
$z $  & 0.084 & Redshift \\
$t_{\rm rise} $  & 1.8--4.0 days & Rise time to peak\tablenotemark{a}\\
$t_{1/2,{\rm rise}} $  & $2.5\pm0.5$ days & Time from half-max to peak\tablenotemark{a} \\
$t_{1/2,{\rm fade}} $ & $8.3\pm1$ days & Time to decay to half-max\tablenotemark{a} \\
$M_{g,{\rm peak}}$ & $-$20.1 & Peak $g$ absolute magnitude \\
$M_{r,{\rm peak}}$ & $-$19.8 & Peak $r$ absolute magnitude \\
$L_{\rm bol, peak}$ & 2$\times$10$^{44}$ erg\,s$^{-1}$& Peak observed UVOIR luminosity \\
$E_{\rm rad}$     & 10$^{50}$ erg & Total UVOIR radiative output \\
$v_{\rm CSM}$    & $-$2200 km\,s$^{-1}$ & Velocity of deepest absorption \\
$v_{\rm max}$    & $-$4500 km\,s$^{-1}$ & Max. blueshift of narrow lines \\
$v_{\rm phot}$    & $-$30000 km\,s$^{-1}$ & Photospheric expansion velocity \\
$M_{\rm *,host}$  & 4.7$\times$10$^{9}$ $M_\odot$ & Host stellar mass \\
SFR$_{\rm host}$  & 0.69 $M_\odot$yr$^{-1}$ & Host star-formation rate \\
12+log[O/H] & $8.35 \pm 0.01$  & Host oxygen abundance   \\
\enddata
\tablenotetext{a}{Times are in the rest-frame and are in the observed $g$-band ($\lambda_{\rm rest} \sim 4500$\,\AA).}
\end{deluxetable}

In the following section we interpret these observations in the context of the progenitor star, its CSM, and the nature of the explosion itself.

\section{Discussion}
\label{sec:discussion}

\subsection{A Highly Chemically-Evolved Progenitor}

The spectra reveal a progenitor star that has lost all of its H, and which is also depleted in He and N.  These properties describe both the narrow (CSM) features and the broad (ejecta) features, and it is clear that the SN represents the explosion of a heavily stripped star into a dense nebula of material recently expelled from its surface.  

An important question is whether the weak He features indicate a qualitatively distinct composition from SNe~Ibn or merely a difference in ionization.  Helium can be a notoriously difficult element to interpret in SN spectra, since non-LTE effects are required for He features to be observable \citep{Li2012,Dessart2012}.  
The almost complete lack of N (alongside that of He) supports the case that the composition is genuinely distinct from that of SNe~Ibn.  In H-burning massive stars, the CNO cycle continuously converts H to He but also converts most existing C and O to N; 
CNO-processed material is expected to have $X_N / X_C \gtrsim 10$ \citep{Gamow1943,Crowther2007}.  In contrast, during the He-burning phase, He is converted to C and O via the triple-alpha process, but N is simultaneously consumed by conversion to Mg and Ne, leaving it heavily depleted.  The absence of detectable N in the UV provides evidence that by the time of explosion virtually the entire remaining star (including its surface, as revealed by the CSM) had experienced triple-alpha processing.

As noted by \cite{GalYam2021}, the velocities and abundance patterns in SNe~Ibn vs. SNe~Icn strongly parallel what is seen in WR (WN vs. WC) stars. 
This does not guarantee that the progenitors \emph{are} WR stars similar to the ones seen in the Milky Way and nearby galaxies; indeed, in \S \ref{sec:massloss} we demonstrate that the properties of the SN~Ibn/Icn progenitor stars shortly before explosion must be quite different from known WR stars.  However, these properties do suggest that the SN~Ibn/Icn progenitors must share two essential characteristics with WR stars: surface abundance patterns from envelope stripping, and high-velocity mass loss.

\subsection{Dramatically Enhanced Pre-explosion Mass Loss}
\label{sec:massloss}

The fast evolutionary timescale SN\,2021csp (a very fast rise, followed by a rapid decline) can only be practically explained by CSM interaction, for reasons detailed in previous works on similarly rapid and luminous objects (e.g., \citealt{Rest2018}): the decline is too fast if radioactive decay of heavy elements is responsible for the heating, but the rise is too slow (and the peak too luminous) to be shock cooling of a supergiant envelope.  Qualitatively, this is consistent with the spectroscopically-inferred notion of a CSM-interacting transient, and indeed our early-time observations provide some of the most direct evidence yet that fast-rising blue transients (of all spectroscopic types) do indeed result from strong CSM interaction.  However, the properties of the CSM are quite extreme for a WR wind.

The SN reaches a peak luminosity of $\sim 2 \times 10^{44}$\,erg\,s$^{-1}$ on a timescale of only 3\ days, and over the course of the first 10\ days (when interaction is the only viable source of energy deposition) the radiative energy release is $\sim 10^{50}$\,erg.  While this is only a few percent of the kinetic-energy budget of a typical SN, substantial CSM is required to decelerate the ejecta over this timescale.

For a SN powered by CSM interaction, the pre-SN mass-loss rate can be related to the observed bolometric luminosity in a simple way assuming basic physical principles (see also \citealt{Smith2017book}).  A star losing mass isotropically at a constant velocity $v_{\rm CSM}$ but potentially variable mass-loss rate $\dot{M}$ will produce a wind nebula with density profile $\rho(r) = \dot{M}/(4 \pi r^2 v_{\rm CSM})$.  
The SN shock then expands into this nebula at a speed
$v_{\rm ej}$, sweeping up matter at a rate $dM/dt = v_{\rm ej} \rho r^2 = v_{\rm ej} \dot{M} / (4 \pi v_{\rm CSM})$.  In the SN shock frame, this matter is suddenly decelerated and its kinetic energy is converted to heat; some fraction $\epsilon$ of this energy is then released as thermal radiation.  Thus, the luminosity is related to the mass-loss rate as

$$L_{\rm bol} = \frac{1}{2} \epsilon \dot{M} \left(\frac{v_{\rm ej}^3}{v_{\rm CSM}}\right). $$

For a variable mass-loss rate, the SN luminosity at  post-explosion time $t$ probes the mass-loss rate at pre-explosion time $-t(v_{\rm ej}/v_{\rm CSM})$.

For SN\,2021csp, we have $v_{\rm CSM} \approx 2000$\,km\,s$^{-1}$ (from early-time spectra), and $v_{\rm ej} \approx 30000$\,km\,s$^{-1}$ (from photospheric modeling).  For these parameters the mass-loss rate is

$$\dot{M} = 0.24 \left(\frac{L}{10^{44}\,{\rm erg\,s^{-1}}}\right) \left(\frac{\epsilon}{0.1}\right)^{-1}\, M_{\odot}\,{\rm yr}^{-1}.$$

Thus, at a time mapping to the bolometric peak of the light curve (+3\ days post-explosion, probing mass loss $-45$\ days pre-explosion), the equivalent mass-loss rate of the star must have been close to 0.5\,$M_\odot$\,yr$^{-1}$.  This is $\sim4$ orders of magnitude higher than what is seen in typical WR stars \citep[e.g.,][]{Barlow1981,Smith2017} --- or, indeed, any stars other than luminous blue variables (LBVs) undergoing giant eruptions.

The narrow lines largely disappear by 16\ days, although we have reason to believe (\S \ref{sec:radioactivetail}) that interaction continues to be the dominant power source of the light curve over the remainder of the evolution of the SN.  Under the simplistic assumptions above, the mass-loss rate 1\,yr prior to explosion was $\sim 0.02$\,$M_\odot\,{\rm yr}^{-1}$ while 3\,yr prior to explosion it was 0.005\,$M_\odot\,{\rm yr}^{-1}$, which is still a factor of 100 greater than for typical WR stars.

Based on this, we conclude that the dense and fast CSM indicated by our spectroscopy originates from a pre-explosion giant eruption rather than a WR wind.  The very close separation in time between this eruption and the explosion ($10^{-4}$ of the lifetime of the WR phase) is unlikely to be a coincidence and suggests that the star was undergoing a period of extreme instability, possibly brought on by late stages of nuclear burning, as has been inferred indirectly from observations of a variety of SNe \citep{Yaron2017,Bruch2021,Strotjohann2021} including at least one SN~Ibn \citep{Foley2007,Pastorello2007}.

This is not in contradiction to the notion that a WR star is responsible for the explosion.  The light curves and spectra of SN\,2021csp show that the interaction phase is very short-lived: once the zone of CSM originating from the pre-explosion eruption has been traversed by the shock, the interaction signatures disappear and the optical luminosity plummets, consistent with the explosion expanding into a more tenuous wind from that point onward.  This behavior is quite different from that of SNe~IIn
(which typically continue to interact with CSM for years) but similar to that of all but a few SNe~Ibn.

\subsection{A Low Radioactive Mass}
\label{sec:radioactivetail}

\begin{figure}
    \centering
    \includegraphics[width=0.472\textwidth]{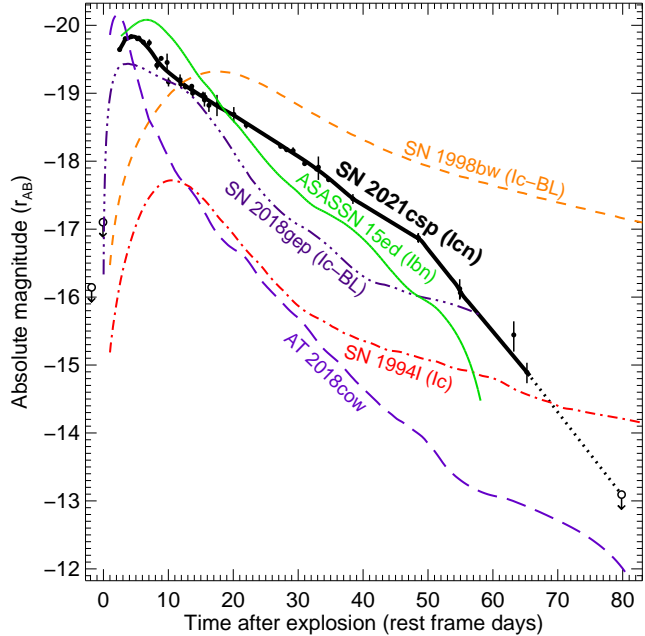}
    \caption{The optical ($r$-band) light curve of SN\,2021csp compared to those of several other transients likely arising from stripped-envelope stars: SN\,1998bw \citep{Patat2001,Clocchiatti+2011}, SN\,2018gep \citep{Ho2019gep}, SN\,1994I \citep{Richmond1996}, ASASSN-15ed \citep{Pastorello2015}, and AT\,2018cow \citep{Perley2019}.  The dotted segment connects the final detection with the deep late-time upper limit.  
    }
    \label{fig:comparelightcurve}
\end{figure}

While the spectra become dominated by broad ejecta features from 15\ days and the luminosity remains high for several weeks thereafter, it is notable that the spectra during this phase do not resemble those of normal SNe~Ib/c: the identifiable features are mostly in emission (not absorption) and the ``temperature'' (a loose concept since the spectra no longer resemble a blackbody) remains high.  Similar behavior is seen in SNe~Ibn, and can be interpreted as the consequence of an inversion of the usual temperature geometry: ejecta are being heated from the shock at the front (producing an emission-dominated spectrum), rather than from radioactive decay from beneath (responsible for the more typical absorption-dominated spectrum).
The distinction from earlier phases is that the optical depth of the pre-shock material has dropped, and the photosphere has receded behind the shock (which may include swept-up CSM material).  

This alone does not rule out the presence of radioactive heating as well: out to 60\ days SN\,2021csp is still quite luminous for a SN and it is easy to imagine a ``typical'' SN~Ib/c explosion buried behind the optically thick shock photosphere --- as is generally presupposed (although rarely demonstrated; \citealt{Pastorello2015}) to exist in SNe~Ibn.  However, the late-time photometric limits provide strong constraints on radioactive heating.

Ordinary (noninteracting, nonsuperluminous) stripped-envelope SNe exhibit two light-curve phases: optically thick and optically thin.  The optically-thick phase is powered primarily by the decay of $^{56}$Ni to $^{56}$Co and manifests as a gradual rise, peak, and decay; the characteristic timescale is set by the diffusion time within the ejecta but is typically about two weeks.
The optically-thin phase is typically powered by the subsequent decay of $^{56}$Co to $^{56}$Fe and follows an exponential curve (linear in time-magnitude space) set by the half-life of $^{56}$Co.  The nickel-heated phase is not constrained by SN\,2021csp, since it is overwhelmed by CSM interaction\footnote{After the initial submission of this manuscript, a separate, independent paper on this event \citep{Fraser2021} was posted, attributing the emission at $t>20$ days to radioactive heating from 0.4 $M_\odot$ of $^{56}$Ni within 2 $M_\odot$ of ejecta (their Figure 9).  This model is ruled out by our deep late-time limits, which fall almost an order of magnitude below the prediction of their model at 80 days ($3\times10^{41}$ erg s$^{-1}$ in their model vs. $L_{\rm bol} < 6\times10^{40}$ erg s$^{-1}$ from our upper limit).}, but the data strongly constrain the presence of a cobalt exponential-decay tail.  Figure~\ref{fig:comparelightcurve} plots the $r$-band light curve of SN\,2021csp versus that of a number of other stripped-envelope SNe, including the well-studied low-luminosity Type Ic SN\,1994I.  The light-curve limit can be seen to fall well below even SN\,1994I at late times, demonstrating that SN\,2021csp was quite ineffective at producing cobalt (and therefore nickel).

Using the empirical method of \cite{Hamuy2003} to convert our late-time $r$-band limit to a constraint on the radioactive mass, we estimate $M_{\rm Ni} < 0.006$\,$M_\odot$, which is lower than what has been inferred for virtually any well-studied SN~Ic to date
\citep{Hamuy2003,Kushnir2015,Anderson2019,Sharon2020,Afsariardchi+2021}.  
This method assumes full gamma-ray trapping, 
which is not a good assumption if the ejecta mass is low. 
To account for this, we employ the gamma-ray trapping prescriptions from \cite{Clocchiatti1997} and \cite{Sollerman1998} to calculate the luminosity at 80\ days for various combinations of $M_{\rm ej}$ and $M_{\rm Ni}$, and compare this with the limiting measurement.  
We have assumed an opacity $\kappa_{\gamma} = 0.03~$cm$^{-2}$~g$^{-1}$, a canonical 
kinetic energy of 10$^{51}$ erg, and a bolometric correction of $M_{\rm bol}-M_{r}=0$, characteristic of late-time stripped-envelope SNe \citep{Lyman2016}.

\begin{figure}
    \centering
    \includegraphics[width=0.472\textwidth]{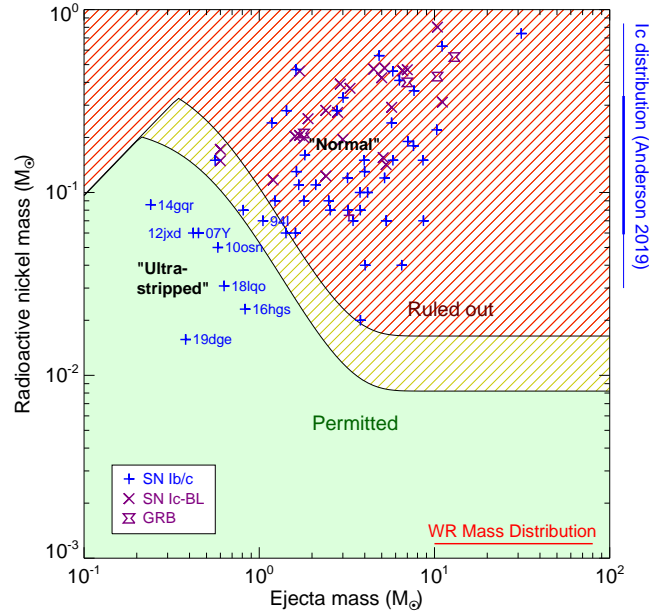}
    \caption{Constraints on the ejecta mass and the amount of radioactive nickel synthesized in the explosion of SN\,2021csp as inferred from the deep late-time NOT observation at 80d (assuming no self-obscuration).  The ``Permitted'' region shows the part of parameter space consistent with this observation; the ``Ruled out'' region shows parameter combinations that would predict optical emission inconsistent with the data (or with $M_{\rm Ni} > M_{\rm ej}$).  The yellow intermediate region is the part of parameter space predicting a flux up to twice that observed, and may be permitted given uncertainties in the models.  Previous Type  Ib/c SNe from the literature are also plotted for comparison.}
    \label{fig:ejectamass}
\end{figure}

The result is plotted in Figure~\ref{fig:ejectamass} (green shading denotes the allowed region).  A strong constraint on the radioactive mass ($M_{\rm Ni} < 0.03$\,$M_\odot$) can be placed even if the ejecta mass is low.  For ejecta masses characteristic of the successful explosion of a WR star, the conditions converge to the full-trapping approximation and the limit is much stronger ($M_{\rm Ni} < 0.008$\,$M_\odot$).

These limits are well below typical values for known Type Ib/c supernovae.  For comparison in Figure~\ref{fig:ejectamass} we plot the estimated ejecta and nickel masses for stripped-envelope SNe from a variety of literature sources (\citealt{Barbarino2021,Anderson2019,Taddia2019,Gagliano2021,Srivstav2014,Stritzinger2009,De2018,De2018a,De2020,Yao2020}, and the compilation of \citealt{Tanaka2009}).  Most such events are within the ruled out region of the diagram, suggesting that SN\,2021csp cannot simply represent a ``normal'' SN Ic exploding into a dense CSM.

A few \emph{atypical} SNe do have very low ejecta masses (and nickel masses) that land within the permitted region.  Events of this nature have sometimes been called ``ultra-stripped'' SNe \citep[e.g.,][]{De2020,Yao2020} due to the need for extremely efficient stripping by a binary companion to explain their origins in terms of a massive star.  SN\,2021csp could originate from such an event, although some other aspects of the explosion are not well-explained in this model: we will discuss this further in \S \ref{sec:ultrastrip}. 

The limits above would be alleviated somewhat if some of the late-time luminosity were obscured by dust produced in the SN shock.
Dust formation has been inferred at late times in at least one SN~Ibn (SN\,2006jc; \citealt{Smith2008,Mattila2008}) and has been appealed to as a partial explanation for the similarly faint late-time emission from that event.   
It is difficult to rule this scenario out entirely, as we lack late-time near-infrared photometry with which to search for dust emission that would be predicted in this scenario.  However, newly-formed dust should not conceal the blue wings of the emission lines (which originate from material at the front of the ejecta). Our spectrum at 88\ days shows no evidence for blueshifted \ion{Ca}{2} emission, suggesting that the line did in fact intrinsically disappear.  More generally, dust formation would have to be extremely rapid (progressing from virtually nonexistent at $\sim 50$\ days to $A_V > 2$\ mag at 80\ days) and the covering fraction would have to be very high ($> 0.9$).  We therefore argue that dust formation is unlikely to explain the late-time rapid fading.

\subsection{Rate Constraints}
\label{sec:rates}

SNe~Icn are clearly rare events: the first example was identified in 2019 \citep[SN 2019hgp, presented by][]{GalYam2021} with only two others discovered thereafter (SN\,2021csp, presented here; SN\,2021ckj, \citealt{Pastorello2021_AN}).  SNe~Ibn are not common either: only 38 are catalogued on the Transient Name Server as of July 2021, compared to 8700 SNe~Ia (which have similar peak luminosities and are detectable to similar distances).  Naively, this suggests that the SN~Ibn rate is 0.4\% of the SN~Ia rate, with SN~Icn rarer by at least a factor of 10.   Given the relative SN~Ia and core-collapse SN (CCSN) volumetric rates \citep[e.g.,][]{Graur2011}, this translates to $\sim 0.1$\% of all CCSNe being of Type Ibn and $\sim 0.01$\% of Type Icn.

This calculation neglects differences in the luminosity function and control times of the various events, as well as any bias in spectroscopic follow-up observations and reporting.  A more robust limit can be calculated from the spectroscopically-complete ZTF Bright Transient Survey \citep{Fremling2020,Perley2020}.  A detailed calculation of the volumetric rates of various CCSN subtypes from BTS will be presented in future work.  For now, we use the methodology from \cite{Perley2020} (including new discoveries through summer 2021) to estimate the SN~Ibn rate for peak absolute magnitudes brighter than $-17.5$ to be 0.1\%--0.5\% of the total CCSN rate.  If we assume that SNe~Icn follow a similar luminosity distribution as SNe~Ibn, the corresponding rate estimate for SNe~Icn is $\sim 0.005$\%--0.05\% of the total CCSN rate.
Regardless of the precise numbers, SNe~Ibn/Icn must be very rare explosions.

\subsection{An Intermediate-Mass Host Galaxy}

\begin{figure}
    \centering
    \includegraphics[width=0.472\textwidth]{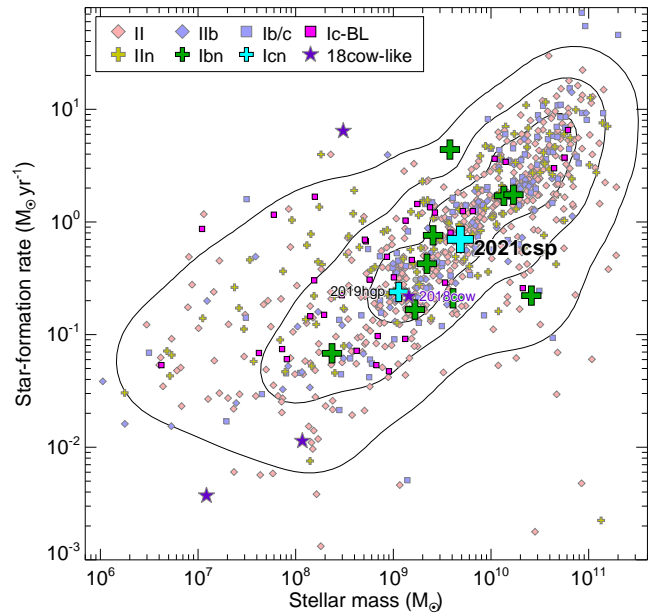}
    \caption{Host-galaxy stellar masses and star-formation rates
    for the iPTF sample of \cite{Schulze2021}, compared to ZTF SN\,Icn and AT\,2018cow-like events.  
    Much like the host of AT\,2018cow, the host galaxy of SN\,2021csp is a generally unremarkable intermediate-mass, star-forming galaxy.  Contours from a kernel-density estimator of the iPTF sample are shown, enclosing approximately [50,80,95]\% of the distribution.
    }
    \label{fig:hostssfr}
\end{figure}

The integrated properties of the host galaxy are similar to those of the Large Magellanic Cloud (LMC) and generally typical of star-forming galaxies.  Figure~\ref{fig:hostssfr} shows basic properties (mass and star-formation rate) compared to a variety of CCSNe from the iPTF survey \citep{Schulze2021}; we have also plotted the host galaxies of all four published AT\,2018cow-like events with radio detections \citep{Perley2019,Coppejans2020,Ho2020koala,Lyman2020,Perley2021} and SN\,2019hgp \citep{GalYam2021}.  The host of SN\,2021csp lies in the middle of the distribution on the star-forming main sequence.  It is also well within the distributions of the hosts of known SNe~Ibn and SNe~Ic-BL.  Thus, for none of these classes is there strong evidence that a highly unusual  progenitor (e.g., extremely metal-poor, ultramassive, or otherwise requiring properties not present within typical massive galaxies) is required.   Much larger samples of SNe~Icn (and SNe~Ibn) will be needed to examine the implications for the nature of the progenitors in detail.

\vspace{0.5cm}

\section{Interpretation}
\label{sec:interpretation}

To summarize, SN\,2021csp represents the explosion of a H/He/N-depleted star into CSM produced by rapid mass loss from the progenitor at very high velocities, likely in the form of an (unobserved) pre-explosion giant eruption.  The explosion itself included very fast ejecta, yet synthesized relatively little radioactive nickel: the SN is dominated at all phases by features of the interaction.  Deep limits at late times rule out a ``classical'' massive, slower-moving component to the ejecta, showing that the explosion did not simply originate from an ordinary class of SN exploding into enhanced CSM.   

Qualitatively similar characteristics were noted for SN\,2019hgp \citep{GalYam2021}, and indeed for many of the prototypical SNe~Ibn as well.  
We consider here three general progenitor scenarios that could explain the observed behavior and its distinction from the general SN~Ib/c population.

\subsection{A Supernova from a Highly Stripped Progenitor?}
\label{sec:ultrastrip}

A variety of faint-and-fast transients in recent years have been interpreted as the results of particularly effective stripping from the binary companion \citep{De2018,McBrien2019,Yao2020}.  In this scenario, late-stage mass transfer is able to effectively remove the large majority of the mass of the progenitor star, leaving behind a core of only a few $M_\odot$ or less \citep{Tauris2013}.  The explosion of such an object naturally produces a SN with limited amounts of ejecta, including radioactive ejecta \citep{Kleiser2018b}.  Should such an explosion occur into a dense surrounding CSM shed by violent pre-explosion instabilities (not naturally predicted in these models, but plausible given the apparent ubiquity of enhanced late-stage mass loss in other SN classes; \citealt{Bruch2021}), the resulting transient would be generally consistent with our observations of SN\,2021csp: as noted in \S\,\ref{sec:radioactivetail}, our late-time limits do not rule out an ``ultra-stripped'' event of this nature.

 Seen in the broader context of the general population of interacting transients, this interpretation is somewhat less satisfying.  Many SNe~Ibn retain significant amounts of hydrogen in addition to helium, and hydrogen is unlikely to persist in the progenitor of a star that has been stripped to a small fraction of its initial mass.  It is possible that the origins of SNe~Ibn and SNe~Icn are distinct, with SNe~Icn representing ultrastripped stars and SNe~Ibn originating from a different mechanism.
However, if this is the case, the strong similarities between the Type Ibn and Type Icn classes (regarding timescale, luminosity, CSM velocity, and late-time behavior) must be ascribed to coincidence.
This, combined with the strong similarities between the spectra of Type Ibn/Icn supernovae and those of (high-mass) WR stars, leads us to consider other potential interpretations.

\subsection{A Pulsational Pair-Instability Eruption?}
\label{sec:ppi}

Another potential explanation for the lack of a late-time radioactive tail is a \emph{nonterminal} eruption that expels only the outer envelope of the star, leaving the remainder intact.  It is already clear from the CSM properties that the star underwent an energetic eruption in the recent past.  If the unstable state that led to that prior eruption subsequently produced a second, higher-velocity eruption, the collision between the two shells could produce a quite luminous transient.  It is unlikely that an ordinary, LBV-style eruption would be sufficient for this, but a more exotic progenitor might produce even more luminous eruptions.  In particular, late-stage pulsational pair-instability (PPI) models have been shown to reasonably reproduce the light curves of SNe~Ibn \citep{Woosley2017,Karamehmetoglu2021}.

While the PPI eruption model cannot be strictly ruled out by any of our observations, we nevertheless disfavor this model for two reasons.
First, the characteristic ejecta velocities associated with pair-instability eruptions are quite low: hundreds to a few thousand km\,s$^{-1}$, much less than what is inferred from the early photospheric expansion rate of SN\,2021csp.  
Second, pair-instability SNe are generally expected to occur primarily or exclusively in extremely metal-poor environments (\citealt{Langer2007,Leung2019}) but the host galaxies of known SNe~Ibn/Icn do not appear to be strongly atypical.  However, this depends on the mass-loss prescription and there may be exceptions \citep{Woosley2017}.

\subsection{Jet Launching from a Failed Explosion of a WR Star?}
\label{sec:fallbackmodel}

The third possibility is that the progenitor of SN\,2021csp (and other SNe~Ibn/Icn) really is a massive Wolf-Rayet star undergoing core collapse, but the SN explosion was extraordinarily weak.

In general, one would expect more-massive progenitors to produce explosions that are both more luminous (owing to the larger cores) and more slowly evolving (owing to the more massive ejecta). There is some evidence that this is the case among ``normal'' SNe~II with identified progenitors \citep{Fraser2011}.  
However, this trend is unlikely to extend to the highest masses: SN simulations suggest that above a certain mass the shock should stall, causing most or all of the star to collapse to form a black hole \citep{OConnor2011,Woosley2015}.
The lowest-luminosity SNe~IIP have sometimes been attributed to marginally successful explosions suffering from substantial fallback \citep{Zampieri2003,Moriya2010}, and it is conceivable that SNe~Ibn/Icn represent equivalent members of the stripped-envelope population \citep{Kleiser2018a}.

The ejecta velocities inferred from the early-time modeling of SN\,2021csp are extremely high, quite unlike what would be expected from a marginally-successful explosion.  The concentration of kinetic energy in a small fraction of the progenitor mass 
could be produced if the explosion is driven by a jet.  There is ample precedent to expect jet formation from WR stars collapsing to form black holes: the original ``collapsar'' model for GRBs in which a rapidly-rotating compact object accelerates ultrarelativistic jets is the most famous \citep{Woosley1993}, but more modest jet energies and velocities can be produced under less-extreme conditions \citep{MacFadyen2001,Piran2019}.  
The interaction between a low-energy jet (or jet cocoon) and a dense shell of inner CSM could lead to a fast, but short-lived, interaction-driven transient of the type seen in SN\,2021csp even as the bulk of the star collapses silently to a black hole.  

Spectropolarimetry (\S \ref{sec:fors2}) does not suggest a highly asymmetric photosphere, representing a potential problem for this model.  Similarly, highly luminous radio emission (another possible jet-interaction signature) is ruled out by our VLA upper limits.  However, the jet in this scenario is generally weaker and slower than in known jet-driven explosions, with the ejecta concealed behind the dense shocked CSM.  There may also be other mechanisms (besides a fallback jet) by which a massive star can produce an incomplete high-velocity explosion.

It should be emphasized that in this model (or in any model), SNe~Ibn/Icn cannot represent the \emph{typical} deaths of WR stars.   Given the abundance of WR stars in the Local Group \citep{Hainich2014,Rosslowe2015} and a lifetime of $10^6$\,yr in this phase \citep{Smith2014review}, the predicted WR death rate is 3--20\% of the CCSN rate \citep{Maoz2010}, at least an order of magnitude in excess of what we inferred in \S \ref{sec:rates}.  
This should not be surprising: the extreme properties inferred from the early-phase observations of SN\,2021csp and similar events require particularly intense pre-explosion mass loss that may in practice be quite rare.
In this scenario, the collapse of a high-mass star would generally produce only a relatively weak transient --- consistent with the lack of good candidates for high-mass progenitors among the general SN~Ib/c population --- but in rare instances (perhaps 1\% of the time) the explosion encounters dense surrounding CSM, leading to a fast-evolving and luminous transient.

It is interesting to note that the one class of successful SNe for which modeling does suggest a significant contribution from high-mass progenitors (SNe~Ic-BL; \citealt{Taddia2019}) has also been connected to jets and engines.  The primary difference is that the much more powerful jets in those events produce more luminous transients and SN explosions and thus do not require dense CSM to be visible.  However, there is increasing evidence that some SNe~Ic-BL do interact with dense surrounding material as well \citep{Corsi2014,Whitesides2017,Chen2018,Ho2020bvc}, raising the possibility of a continuum of WR collapse transients, with the vast range in observable properties explained by variations in the jet power, pre-explosion mass-loss history, and degree of progenitor stripping.

If this is the correct model, it would shed light on the even rarer, even faster-evolving transient population of AT\,2018cow-like transients, which show a number of similarities to SNe~Ibn/Icn  \citep{Fox2019}.  AT\,2018cow and its analogs have also been hypothesized \citep{Perley2019,Margutti2019} to originate from ``failed'' collapses based on some of the arguments presented above: the luminous early transient implies a very fast-moving early component, yet late-time observations provide deep limits on nickel production from the associated SN, demonstrating that they cannot simply represent normal (or even rare) SNe exploding into an unusually dense medium.  AT\,2018cow-like transients show major differences from the SN~Ibn/Icn population, including a complete lack of early interaction signatures and a radio/X-ray ``afterglow'' that is more luminous than the limits on SNe~Ibn/Icn by many orders of magnitude \citep{Ho2019cow,Ho2020koala}.  This difference may be explicable in terms of the relative power and velocity of the jet and the precise geometry of the CSM, or it may be more fundamental.

\section{Conclusions}
\label{sec:conclusions}

The Type Icn SN\,2021csp is one of the most extreme known examples of an interaction-powered fast and luminous transient and also among the best observed.  Its properties, alongside those of SN\,2019hgp (the first SN~Icn) and the general population of SNe~Ibn, provide a challenge to the basic picture of interaction-driven SNe as resulting from the explosions of otherwise ordinary SNe into dense CSM.  The expansion speeds inferred from modeling the rising light curve are much higher than seen in ordinary stripped-envelope SNe, while the late-time flux is too faint for an explosion that produces significant ejecta and/or leads to significant radioactive nucleosynthesis (absent rapid dust formation).  

The properties of SN\,2021csp and other interacting SNe can be explained by a variety of potential models on an individual basis: a moderately-massive star highly stripped by a binary companion, or a nonterminal late-stage pair-instability eruption from an extremely massive star.  However, we argue that the collective properties of this class are best explained under a scenario in which most SNe~Ibn/Icn are produced by partially-successful explosions following the collapse of massive Wolf-Rayet stars.
Specifically, we propose a model in which the direct collapse of a WR star to a black hole launches a subrelativistic jet that interacts with dense CSM shed by the progenitor shortly before explosion.   Given that other stripped-envelope SNe now appear to originate from stars with masses too low to have classical WR progenitors, this raises the possibility that SNe~Ibn/Icn, and engine-driven explosions like SNe~Ic-BL and long-duration gamma-ray bursts, may actually represent the \emph{only} currently-observable manifestation of the collapse of the most massive stars. 

Further studies will be necessary to robustly establish whether this scenario is indeed the correct one.
SN\,2021csp represents one of only two published SNe~Icn, fewer than five spectroscopically-confirmed AT\,2018cow-like events are known, and even the SN~Ibn population is only crudely mapped out (with sparingly few pre-maximum-brightness detections or deep late-time limits).   Fortunately, with ZTF and a number of other wide-area surveys fully operating and with increasing community interest in the fastest transients, the sample is destined to grow (albeit slowly) in the coming years.

The Legacy Survey of Space and Time (LSST; \citealt{Ivezic2019}) with the Rubin Observatory will also play a vital role in this effort.  While the slow cadence of the primary survey is poorly suited to the discovery of fast-evolving transients, the host-galaxy photometric redshifts will make it much more straightforward to distinguish luminous phenomena in high-cadence shallower surveys, providing important synergy with the fast wide-field surveys of the future.  Meanwhile, repeated deep LSST imaging of nearby galaxies may be able to test whether WR stars disappear without a trace, better seek and identify pre-explosion progenitor eruptions in future SNe~Ibn/Icn (and other SNe), and search for low-luminosity transients associated with black-hole fallback even in the absence of strong CSM interaction.\\\

\vspace{1cm}

\textit{Acknowledgments---} We thank Ori Fox, D. Alexander Kann, and the anonymous referee for helpful comments and suggestions.
Based on observations obtained with the Samuel Oschin 48-inch telescope and the 60-inch telescope at the Palomar Observatory as part of the Zwicky Transient Facility project. ZTF is supported by the National Science Foundation (NSF) under grant AST-2034437 and a collaboration including Caltech, IPAC, the Weizmann Institute for Science, the Oskar Klein Center (OKC) at Stockholm University, the University of Maryland, Deutsches Elektronen-Synchrotron and Humboldt University, the TANGO Consortium of Taiwan, the University of Wisconsin at Milwaukee, Trinity College Dublin, Lawrence Livermore National Laboratories, and IN2P3, France. Operations are conducted by COO, IPAC, and UW. The SED Machine is based upon work supported by the NSF under grant 1106171.  The ZTF forced-photometry service was funded by Heising-Simons Foundation grant \#12540303 (PI Graham).
The Liverpool Telescope is operated on the island of La Palma by Liverpool John Moores University in the Spanish Observatorio del Roque de los Muchachos of the Instituto de Astrofisica de Canarias with financial support from the UK Science and Technology Facilities Council. 
Partly based on observations made with the Nordic Optical Telescope, owned in collaboration by the University of Turku and Aarhus University, and operated jointly by Aarhus University, the University of Turku, and the University of Oslo (respectively representing Denmark, Finland, and Norway), the University of Iceland, and Stockholm University at the Observatorio del Roque de los Muchachos, La Palma, Spain, of the Instituto de Astrofisica de Canarias. These data were obtained with ALFOSC, which is provided by the Instituto de Astrofisica de Andalucia (IAA) under a joint agreement with the University of Copenhagen and NOT.
Based in part on observations collected at the European Organisation for Astronomical Research in the Southern Hemisphere under ESO programme(s) 106.21U2 and 106.216C.
Some of the data presented herein were obtained at  the W. M. Keck Observatory, which is operated as a  scientific partnership among the California Institute of    Technology, the University of California, and NASA; the observatory was made possible by the generous financial support of the W. M. Keck Foundation. 
A major upgrade of the Kast spectrograph on the Shane 3~m telescope at Lick Observatory was made possible through generous gifts from the Heising-Simons Foundation as well as William and Marina Kast. Research at Lick Observatory is partially supported by a generous gift from Google. 
We thank the staffs of the various observatories where data were obtained for their assistance.

J.S., S.S., and E.K. acknowledge support from the G.R.E.A.T. research environment funded by {\em Vetenskapsr\aa det}, the Swedish Research Council, under project \#2016-06012. The OKC's participation in ZTF was made possible by the K.A.W. foundation.  E.K. also acknowledges support from The Wenner-Gren Foundation.
M.M.K. acknowledges generous support from the David and Lucille Packard Foundation. 
P.R. has received support from the from the European Research Council (ERC) under the European Union's Horizon 2020 research and innovation programme (grant agreement 759194 -- USNAC).
The research of Y.Y. is supported through a Benoziyo Prize Postdoctoral Fellowship and a Bengier-Winslow-Robertson Fellowship.
T.-W.C. acknowledges EU funding under Marie Sk\l{}odowska-Curie grant H2020-MSCA-IF-2018-842471.
R.L. acknowledges support from a Marie Sk\l{}odowska-Curie Individual Fellowship within the Horizon 2020 European Union (EU) Framework Programme for Research and Innovation (H2020-MSCA-IF-2017-794467).
The GROWTH Marshal \citep{Kasliwal2019} development was supported by the GROWTH project funded by the NSF under grant 1545949.
A.V.F.'s group is supported by the Christopher R. Redlich Fund, the Miller Institute for Basic Research in Science (where A.V.F. is a Miller Senior Fellow), and many individual donors.
A.G.Y.’s research is supported by the EU via ERC grant No. 725161, the ISF GW excellence center, an IMOS space infrastructure grant and BSF/Transformative and GIF grants, as well as The Benoziyo Endowment Fund for the Advancement of Science, the Deloro Institute for Advanced Research in Space and Optics, The Veronika A. Rabl Physics Discretionary Fund, Minerva, Yeda-Sela and the Schwartz/Reisman Collaborative Science Program;  A.G.Y. is the incumbent of the The Arlyn Imberman Professorial Chair.

{IRAF} is distributed by the National Optical Astronomy 
Observatories, which are operated by the Association of Universities for 
Research in Astronomy, Inc., under cooperative agreement with the NSF.


\vspace{5mm}
\facilities{HST(STIS,COS), Swift(XRT,UVOT), PO:1.2m(ZTF), PO:1.5m(SEDM), Hale(DBSP), LT(IO:O,SPRAT), VLA, Gemini:Gillett, NOT, VLT:Antu (FORS2), Shane (Kast), Keck:I (LRIS)}

\software{IRAF \citep{Tody1986}, HEASoft \citep{Blackburn1995}, Pypeit \citep{Prochaska2020}, LPipe \citep{Perley2019b}, DBSP-DRP \citep{dbsp_drp2021}, PyNOT \citep{pynot}, Galfit \citep{Peng2002,Peng2010}}



\end{document}